\documentclass[preprint]{aastex}

\def\simgr{\,\hbox{\hbox{$ > $}\kern -0.8em \lower 1.0ex\hbox{$\sim$}}\,}
\def\simle{\,\hbox{\hbox{$ < $}\kern -0.8em \lower 1.0ex\hbox{$\sim$}}\,}

\shortauthors{THORSTENSEN.}
\shorttitle{Parallaxes of Cataclysmics}
\begin{document}
\title{Parallaxes and Distance Estimates for Fourteen Cataclysmic Variable Stars
\footnote{Based on observations obtained at the MDM Observatory, operated by
Dartmouth College, Columbia University, Ohio State University, and
the University of Michigan.}
}

\author{John R. Thorstensen}
\affil{Department of Physics and Astronomy\\
6127 Wilder Laboratory, Dartmouth College\\
Hanover, NH 03755-3528;\\
john.thorstensen@dartmouth.edu}
% \author{Joseph O. Patterson, Jonathan Kemp\altaffilmark{2}, Jules Halpern}
% \affil{Department of Astronomy, Columbia University\\
% 538 West 120th Street, New York, NY 10027;\\
% jop@astro.columbia.edu, j.kemp@jach.hawaii.edu,jules@astro.columbia.edu}
% \altaffiltext{2}{Also at Joint Astronomy Center, Hilo, Hawaii.}

\begin{abstract}
I used the 2.4 m Hiltner telescope at MDM Observatory
in an attempt to measure trigonometric parallaxes for 14 
cataclysmic variable stars.  Techniques are described in detail.
In the best cases the parallax uncertainties are below 1 mas,   
and significant parallaxes are found for most of the program
stars.  A Bayesian method which combines
the parallaxes together with proper motions 
and absolute magnitude constraints is developed and 
used to derive distance estimates and confidence intervals.
The most precise distance derived here is for WZ Sge, 
for which I find $43.3 (+1.6, -1.5)$ pc.
Six Luyten Half-Second stars with previous precise 
parallax measurements were re-measured to test the techniques,
and good agreement is found.  

\end{abstract}

\section{Introduction}

Cataclysmic variables (CVs) are binary stars in which a white dwarf accretes
matter from a close companion, which usually resembles a lower-main-sequence
star.  The dwarf novae are a subclass of CVs which show distinctive outbursts,
thought to result from an instability in an accretion disk about the white
dwarf.  The AM Herculis stars (sometimes called ``polars'') are another type of
CV, in which the accreted material is entrained in a strong magnetic field
anchored in the white dwarf, forming an accretion funnel above the magnetic
poles.  \citet{warn} has written an excellent monograph on CVs.  

Distances for various types of CVs are fundamentally important to physical
models 
(see, e.g., \citealt{beuermanneferi}), but they have not been easy to obtain
\citep{berriman87}.  Historically, none were near enough for
trigonometric parallax.  \citet{kamper79} published parallaxes for some of the
brightest dwarf novae, but these have proven to be incorrect
\citep{harrison99}.  The {\it Hipparcos} satellite obtained useful parallaxes 
only for a handful of the apparently brightest CVs \citep{duerbeck99}.
\citet{kraftluyten65} used statistical parallaxes (proper motions and radial
velocities) to determine rough absolute magnitudes for dwarf novae at minimum
light.  When the secondary star is visible in the spectrum (which it often
isn't), and the orbital period $P_{\rm orb}$, is known, it is possible to 
estimate a distance
spectroscopically, by combining the surface brightness vs. $T_{\rm eff}$
relation with constraints on the secondary's radius from the Roche geometry.
\citet{bailey81} employed a variant on this method which used $K$-band
photometry.  Because $K$-band surface brightness depends weakly on spectral
type, and at a known $P_{\rm orb}$ the Roche tightly lobe constrains the secondary's
radius, the apparent $K$ magnitude can be used to set a lower limit on a CV's
distance under the assumption that {\it all} the $K$-band light arises from the
secondary.  \citet{shm}, among others, employed this technique.  But a lower
limit is not a distance, and the complex nature of the disk's emission makes the 
method problematic in many cases \citep{bsc}.  Finally, for a
small number of systems distances have been derived from spectra of the exposed
white dwarf during intervals of low accretion (see, e.g., \citealt{sionwz}).

Only recently have a few accurate parallaxes of CVs become
available from the Fine Guidance Sensor (FGS) on HST.  \citet{harrison99},
in a breakthrough paper, reported the first accurate distances for
the bright dwarf novae SS Cyg, U Gem, and SS Aur, and from these
found that the secondaries of these dwarf novae are somewhat
too luminous for the main sequence \citep{harrison00}.  FGS
parallaxes are also available for the novalike variables  
RW Tri \citep{mcarthurrwtri} and TV Col \citep{mcarthurtvcol}.  
And just as the present paper was being completed, T. Harrison
kindly communicated a draft of \citet{harrison03a}, with precise
parallaxes for WZ Sge and YZ Cnc, two of the objects studied
here. 

Meanwhile, the precision of ground based parallaxes has 
advanced dramatically thanks to the advent of CCD detectors.
\citet{monetdahn83} described early CCD parallax work and 
showed that very high precision was possible; \citet{monet92}
(hereafter USNO92) gave detailed descriptions of their
procedures and accurate parallaxes for dozens of 
stars, mostly from the Luyten Half-Second (LHS) list.  Many of the
parallaxes in USNO92 have formal uncertainties less than 
1 mas (= $10^{-3}$ arcsec), which is not much worse than
HST parallaxes.  \citet{dahn02} give more recent results
from the USNO program and discuss further refinements of their
technique.  Only a handful of active CCD parallax programs
exist, so the present project was partly motivated by 
curiosity as to how accurately a non-specialist could
measure parallaxes using a general-purpose telescope.

While HST parallaxes have unsurpassed precision, HST observing 
time is limited and relatively few objects can be observed.  
I therefore attempted ground-based CCD parallax
determinations of a sample of cataclysmic variable stars.
The sample is not meant to be complete or representative, 
but was selected informally on the basis of brightness, perceived
likelihood of a positive result, astrophysical interest,
and observational constraints.  This paper describes this program as follows.
Because this is a new program, sections 2 and 3 describe the 
observations and analysis procedures in detail.
Section 4 discusses the Bayesian method used
to estimate distances from the parallaxes, proper motions,
and magnitudes.  Section 5 presents the results for the
individual CVs.  A brief discussion follows in Section 6,
and Section 7 offers conclusions.

\section{Observing Procedures and Data Reduction}

All the observations are from the {\it f}7.5 focus of the Hiltner
2.4 m telescope at MDM Observatory on Kitt Peak, Arizona. 
A thinned SITe 2048$^2$ CCD yielded a scale of $0''.275$
per 24 $\mu$m pixel.  The 50 mm square filter vignetted the
field of view, so the images recorded were $1760^2$.  
The same detector was used throughout the program.

Table 1 lists the targets and gives a journal of the 
observations.  Because the telescope is `classically'
scheduled in observing runs, the observations come in short
runs of a few days separated by months or years.  Thus
even the best-observed targets have observations at
only a modest number of independent epochs.  It was not
practical to schedule runs 
to maximize the parallax displacements of individual objects.

USNO92 explain {\it differential color
refraction} (DCR).  When different spectral energy distributions
are convolved with a finite passband, different effective wavelengths
result, which suffer different amounts of refraction because
of atmospheric dispersion.
This effect was severe for USNO92 because they used a 
wide passband.  To minimize DCR, I chose
a filter approximating Kron-Cousins $I$, which 
is narrower than the USNO92 filter and farther to the red,
where atmospheric dispersion is reduced.
All the parallax observations were taken with this filter.
Most exposures were taken within $\pm 1$ hr of the 
meridian, in order to further minimize DCR effects and
other problems which might arise from telescope flexure and
such.  Because the DCR effects were much less severe than
for the USNO, some exposures from larger hour angles were 
included.

On the first run, the CCD was inadvertently aligned with
its columns about 2.4 degrees from true north.  This
(mis)alignment was maintained throughout the program, with the
exception of 1999 June, when the alignment was not set correctly.
There was no obvious problem with the 1999 June data, so this step
may not have been necessary.

When each target was first observed, it was centered approximately
on the CCD and its pixel coordinates noted.  For all subsequent
observations, the pointing was reproduced within roughly 20 pixels
in order to minimize the effects of any optical distortion
and to allow use of a consistent set of reference stars.

In order to avoid saturating the reference or (sometimes) program star images,
exposures were generally kept to $< 100$ s, reaching 25 s in some cases.
Reading and preparing the CCD took more than two minutes,
so efficiency suffered.  Many (typically $\sim 10$)
exposures were taken at each pointing.  From time to time the telescope 
was `dithered' by a few pixels between exposures in a sequence.  

As one might expect, the seeing strongly affected the results.  Pictures taken
in poor seeing had large fit residuals.  For this reason, few data
are included from images with seeing worse than 1.5 arcsec FWHM, and the
majority of the data are from images with $<$ 1 arcsec seeing.   The very best
images included are around $0.6$ arcsec, still not quite undersampled.  

When the sky was suitably cloudless, exposures in the $UBVRI$ filters (or
sometimes only $V$ and $I$) were added to the program, together with
standard star fields from \citet{landolt92} to allow
transformation to standard magnitudes.  Photometric
exposures were obtained on at least two nights, and the results were averaged
after analysis.  The consistency was generally better than 0.05 mag.

The data were reduced using standard IRAF\footnote{The Image Reduction and
Analysis Facility, distributed by the National Optical Astronomy Observatories.}
routines for bias subtraction and
flat field correction.  The flat fields were constructed from offset, medianed
images of the twilight sky.

\section{Data Analysis}

Star centers were measured with the IRAF
implementation of DAOPHOT (originally written by \citealt{stetsondao}), which
constructs a model point-spread function (PSF) from selected stars and fits
these to the program stars.   Because the centroid information is contained in
the steep sides of the PSF, a small fitting radius was used, generally 0.8
arcsec.  For some of the later measurements, the fitting radius was adapted to
the seeing on the individual pictures.

The measurement procedure was automated as follows.  First, the average of
several of the best pictures (the `fiducial' frame) was examined to select a
set of stars to measure and a set of suitable PSF stars.  Next, lists of stars
on all the pictures frames were generated using {\it daofind} or {\it
SExtractor} \citep{bertin96}.  A computer program matched objects on these
lists to the corresponding objects on the fiducial frame, and the matches were
used to transform the program and PSF star coordinates to the system of each picture.
The DAOPHOT measurements proceeded automatically.

To determine the true scale and orientation of the fiducial frame, the star
images were matched to the USNO A2.0 catalog \citep{mon96}, which is aligned
with the ICRS (essentially J2000).  In most fields several dozen stars were
matched, with plate solutions typically having RMS residuals of $0.''3$,
mostly from the centering uncertainty of the USNO A2.0 and 
proper motions since the USNO A2.0 plate epoch.  Given
the number of stars in the solutions and the size of the field, the image
scales derived from these fits should be accurate to a few parts in $10^4$, and
the orientation should be accurate to a $\sim 0.03$ degree.  
Using the scale and orientation, the pixel
coordinates of the fiducial stars were transformed to tangent plane coordinates
$X_{\rm fid}$ and $Y_{\rm fid}$, with the program object at the origin. These
coordinates correspond closely to $\Delta \alpha$ and $\Delta \delta$ over a
small field.

Once the fiducial stars were characterized, a computer program collated the
DAOPHOT image centers from the original pictures, and generated a
master raw data file containing the fixed information about the
measured stars, the celestial location, the Julian dates of the exposures, and
the measured image centers $(x_{\rm DAO}, y_{\rm DAO})$  from all the pictures.  In addition, a weight
of zero or one was assigned to each star indicating whether it was to be
used in generating coordinate transformations.  The program star was never used
for the coordinate transformations, and other stars were eliminated from the
transformations if preliminary analysis showed large scatter or (in some
cases) large proper motions or parallaxes.

The analysis proceeded in several steps as follows:  

{\it 1: Computable corrections.} For each star, corrections for differential
refraction, differential aberration, and DCR were computed, in the ($X_{\rm
fid}, Y_{\rm fid}$) system.  A transformation was derived between 
($x_{\rm DAO}, y_{\rm DAO}$) and ($X_{\rm fid}, Y_{\rm
fid}$).  The net correction was transformed back to the ($x_{\rm DAO}, y_{\rm
DAO}$) system, and added to the original coordinates.  Thus the coordinates
were `born corrected'.  Routines adapted from {\it skycalc}\footnote{This
time-and-the-sky program was written by the author, and is available from
ftp.iraf.edu in the contrib directory.} performed the spherical trigonometry
calculations.  Tests showed that with the exception of DCR (discussed below), 
these corrections generally made relatively
little difference to the results, because their effects were largely
absorbed by the `plate model' later in the process.

The DCR correction calls for some discussion.  Early
experiments with fields deliberately taken both near and far from the meridian 
suggested a DCR coefficient near 7 mas per unit ($\tan z$) per unit $(V - I)$,
whereas the polynomial given by USNO92 implies a value of 29 in the same units
for their broader passband.
I checked the empirically-derived DCR coefficient using a procedure outlined by
\citet{gublertytler}, as follows.  Library spectra from \citet{pickles} were
convolved with passbands from \citet{bessell} to compute effective wavelengths
as a function of $V - I$, and a {\it slalib} \citep{wallace} routine was used
to to find the refraction as a function of wavelength.  The final result was 5
mas per unit ($\tan z$) per unit $(V-I)$,
in reasonable agreement with the empirical 7.  To verify the procedure,
the calculation was repeated for the USNO92 passband (approximated as flat
across their coverage), and their value of 29 units was recovered successfully.  

{\it 2: Position averaging.} The ($x_{\rm DAO}, y_{\rm DAO}$) coordinates were
transformed to the system outlined by ($X_{\rm fid},Y_{\rm fid}$), using
a four-constant plate model (which allows only shifts in zero point, a rigid
rotation, and a scale change).  These positions were averaged to create refined
positions ($X_{\rm fid2}, Y_{\rm fid2}$) for each star.  The errors in these
positions were much reduced because of averaging over many frames and because
of the previous step's removal of computable offsets.

{\it 3: Iteration.} The transformations between ($x_{\rm DAO}, y_{\rm DAO}$)
and the $XY$ system were computed again using ($X_{\rm fid2}, Y_{\rm fid2}$) as
the target coordinates and using a more flexible plate model of the form 
$$X' = a_0 + a_1 X + a_2 Y + a_3 X^2 + a_4 XY + a_5 Y^2 + a_6 X  r^2 + a_7 Y r^2,$$
where $r$ is the radial distance from the fiducial point at the middle
of the field,
and similarly for $Y$.  This model was needed to adequately account for
variable systematic distortions across the wide field of view.  The
($x_{\rm DAO}, y_{\rm DAO}$) coordinates were transformed onto this system, and
residuals were formed by subtracting away ($X_{\rm fid2}, Y_{\rm fid2}$).
These formed the time series of offsets in $X$ and $Y$ to be fitted in the next
step.

{\it 4: First pass fitting.} The residuals of each star were fitted to $$X(t) =
X_0 + \mu_X (t - \bar t) + \pi p_X(t)$$ and similarly for $Y$,
% $$Y(t) = Y_0 + \mu_Y (t - \bar t) + \pi p_Y(t),$$
where $t$ is the time (Julian date) and $\bar t$ is its mean, $X_0$ and $Y_0$
are small offsets relative to the star's adjusted fiducial position, $\mu_X$
and $\mu_Y$ are proper motions, and $p_X(t)$ and $p_Y(t)$ are the parallax
factors along each axis at time $t$ for the star's $\alpha$ and $\delta$, which
were computed using a {\it skycalc} routine.  The fitting was done in a
somewhat unorthodox manner; first estimates of the parameters were computed
using linear least-squares fits to $X$ and $Y$ separately, and then a numerical
steepest descent algorithm was used to minimize $$\sum_i \left\{[X_i - X(t)]^2
+ [Y_i - Y(t)]^2\right\},$$ where $(X_i, Y_i)$ is the $i$-th data point; this
explicitly couples the $X$ and $Y$ solutions through the common parameter
$\pi$.   Fig.~1 gives an example of how the residuals are fitted 
(though for the data shown all the iterations described below have
been performed).

{\it 5: Iteration (again).}  The coordinate transformations
were computed again, this time adjusting the 
stars' fiducial positions for the proper motion and
parallax displacements at the epoch of each picture.  
In practice this made little
difference, since stars with large enough motions to
matter were generally eliminated from the fit
earlier in the process because of their large residuals.
Residuals between the stars' positions on individual
frames and their mean positions were again computed
and used as the basis for the final step.

{\it 6: Final fitting.} The residuals were fitted again,
as in step (4) above.  Formal uncertainties were estimated using
the procedure outlined by \citet{cash79}; essentially,
error bars were drawn at critical levels of the mean
square fit residual.  This procedure assumes that the 
fitted parameters are uncorrelated.  This is only 
defensible in this case if the observations extend over
enough epochs to cleanly separate parallax and proper
motion.  The observations reported here generally satisfy
this criterion.

{\it 7: Human Editing.}  The reduction code allows the user
to examine and edit the input data; the whole process
(1 -- 6) can then be run again.  The fit residuals from
individual pictures were examined and pictures with
particularly large scatter, generally due to poor seeing,
were removed.  The comparison star fits were reviewed
individually, and objects with large scatter (due to
faintness or other difficulties such as incipient duplicity)
were eliminated from the reference grid.  The reference
stars making the final cut generally had RMS residuals
below 10 mas.  In some instances reference stars were
eliminated because of their large proper motions, which
would skew the zero point of the proper motions.

The result of steps 1 -- 7 was a set of parallaxes, proper motions, and formal
errors for all the stars which had been measured in each field.  
Table 2 (available in full in the electronic version of this paper)
presents this information, along with the celestial coordinates,
$V$ and $V-I$ magnitudes, RMS residuals of the fits, and statistical
weights.  

A correction to absolute parallax was estimated as follows.  For each
star used for the reference frame, a distance was estimated from
the measured $I$ and $V-I$ color, using typical main-sequence values
tabulated by \citet{pickles}.  The straight mean of the estimated
reference-star parallaxes was used to correct the relative parallax to
absolute.  Reddening of the reference stars was not taken into account,
nor was the possibility that they might be giants, which we cannot
exclude.  At high latitudes, giants of the apparent magnitude of the
reference stars would be far out in the halo and hence unlikely {\it a
priori}; at lower latitudes, where many more reference stars were
available, a few mis-identified giants would lead to a slight
overestimation of the very small correction.  The use of $I$ magnitudes
mitigated the effects of absorption to some extent; in any case,
unaccounted-for extinction would somewhat counter-intuitively tend to
make the stars appear closer than their true distances, because the
reddening would make the stars appear later-type, and hence absolutely
fainter, and hence closer than their true distance.  The extinction
would also make the stars appear fainter (and hence farther away), but
the former effect more than compensates for the latter.  Accordingly,
this estimate is in effect an upper limit to the correction.  The
corrections were in all cases small, of order 1 mas, and the uncertainty
in the correction was ignored. 

The number of stars measured in each field was large enough to allow an
alternate calculation of the parallax error, as follows.  A set of stars
was chosen for proximity on the sky and similarity in brightness, and the
scatter of the fitted parallaxes of these stars was taken as an alternate
parallax uncertainty.  The magnitude and radius window was adjusted
to include $\sim 10$ stars or more in the sample; typically stars within
3 arcmin and $\pm 1$ mag of the program star were included, but this
varied widely based on how many stars were available.   The scatter
was computed both around zero and around the stars' photometric parallaxes,
but the photometric parallax adjustments were small enough, and the
errors large enough, that this made little difference in practice.  
This measure of parallax uncertainty was usually somewhat greater than the
fit errors above, but they were not dramatically larger, indicating that the
fit errors were not too far off.  Nonetheless, these measures are
probably more faithful indicators of the true external error, and
they were used in the distance estimates given later, except in those
cases in which the 
estimated external error was {\it less} than the formal error, which
could happen because of the small number of stars involved.

{\it Error of a single measurement.} 
The scatter around the parallax and proper motion
fits for the best, stablest comparison stars is around 6 mas
(vector rms error).  This is about double the error
obtained at USNO with a similar CCD and slightly poorer
image scale \citep{dahn02}.  

The short exposure times used in the present study 
may contribute to this in the following way.  
The effectiveness of `tip-tilt' adaptive optics schemes
demonstrates that bulk image motion is a major 
contributor to seeing.  Typical amplitudes of the
bulk image motion are 200 mas or so.  The coherence
time of the atmosphere in good seeing is of order 30 milliseconds,
so one obtains effectively 30 independent samples
of the image motion for each second of exposure time,
or 1800 such samples in a typical 60 s exposure.  
The image centers should therefore be displaced
by approximately (200 mas) /$\sqrt{1800}$, or around
5 mas. If all the stars suffered the same displacement
this would be of no concern, but the seeing only
correlates over the isoplanatic patch, which is typically
less than a couple of arcmin in size.  Indeed, examining
residual maps from individual frames in succession, there
is a strong impression that residuals correlate over 
patches of roughly this size and vary from picture to picture.

The USNO telescope is also a purpose-built astrometric telescope,
while the Hiltner telescope is not.  Unmodeled optical distortions
over the relatively wide field may contribute to the error budget;
the fairly complicated plate model needed to adequately model the
reference frame indicates that this is likely the case.

{\it Accuracy of the Proper Motions.}  The formal errors
in the proper motions are generally very small, of order
1 mas yr$^{-1}$ in most cases.  However, no attempt
was made to put these in an absolute frame.  Scatter
diagrams of the proper motions in all of the fields 
suggest that small-number statistics typically affect the mean 
reference star motion at the 2 to 3 mas yr$^{-1}$ level;
as noted earlier, exceptionally high proper motion reference
stars were eliminated to avoid skewing the zero point 
excessively.  At high latitudes the reference star grids were relatively
sparse and the proper motions noticeably larger than at
low latitude, where more distant reference stars are 
available; both these effects make the proper motion zero
point at high latitude somewhat more loosely defined than
at low latitude.  The high-latitude frames often have
detectable galaxies, which could in principle constrain the
zero point, but it was felt that the centroids of these
extended objects could not be defined with sufficient
precision to make this worthwhile.

%\subsection{Internal consistency}.  
%The measurements of other field stars can be used for 
%an independent estimate of the uncertainty.  For each field,
%a sample of at least xxx six check stars was selected, all of which 
%lay within a radius around the program object's position 
%and which were within a magnitude window centered on the program object's
%magnitude.  The radius and window used to define the sample
%were adjustable; in sparse fields the radius and window
%were made larger to include enough stars.  The rms scatter
%of these parallaxes gave a conservative measure of the external
%error, usually somewhat larger than the formal error estimate from
%the fits.  The estimate was refined somewhat by adjusting each
%field star parallax to absolute using the correction above, 
%and then comparing it to the star's photometric parallax
%based on its $V-I$ color and $I$ magnitude (see above); the
%scatter in these measures was generally slightly smaller than the
%scatter of the apparent parallaxes around zero, but the adjustments
%difference was generally not significant.  

\subsection{Checks Using Nearby Stars}

To check the above procedures, I re-observed several of the Luyten
Half-Second (LHS) stars with precise parallaxes published by
\citet{monet92}.  These were for the most part not observed as
extensively as the program stars, but the agreement is nonetheless
satisfactory.  Because similar sets of comparison stars were used in the
two determinations, the correction to absolute parallax is ignored, as it
should have almost no effect on the comparison.

Table 3 lists the results.
A few of these deserve comment.  LHS 429 is extremely red, and hence
requires a relatively large DCR correction.  Turning off the DCR
correction decreases $\pi_{\rm rel}$ by about 10 mas, ruining the good
agreement with \citet{monet92}, and offering a somewhat roundabout check
on the DCR correction procedure.  LHS 1801 and 1802 are a common proper
motion pair, for which the USNO parallaxes differ by 2.5 of their mutual
standard deviation; the MDM parallaxes are internally consistent, though
slightly smaller than the USNO parallax.  Interestingly, there are
slight but significant differences in the proper motions between the two
stars which are nearly identical in the two determinations, suggesting
that orbital motion is detected.  Finally, a foreground L3.5 dwarf was
discovered serendipitously among the stars measured in the field of LHS
1889 \citep{thorkirk03}.

\section{Distance Estimation}

For a uniform distribution of objects in space, the distribution of true
parallaxes $\pi$ is proportional to $1/\pi^4$, because the volume
element at a distance $r = 1/\pi$ is proportional to $r^2\, dr$.  From
this, \citet{lutzkelker} show that if parallaxes have substantial
relative uncertainties, they tend to systematically underestimate
distance.  If the measured parallax $\pi_0$ is assumed to have
Gaussian-distributed measurement errors with standard deviation
$\sigma_\pi$, $\pi$ is distributed as $$L(\pi | \pi_0) = {e^{-(\pi -
\pi_0)^2 / 2 \sigma_\pi^2} \over \pi^4}.$$ (The vertical bar in the
argument of $L$ is read as `given', as is standard; this is therefore
`the likelihood of $\pi$ given a measurement $\pi_0$'.) This expression
does not include any {\it a priori} constraint on the distance -- the
objects can be any absolute magnitude, for example.  In their treatment
the local maximum of $L$ is taken as the best estimate of the true
parallax.  If the relative error $\sigma_\pi /\pi_0$ is small, the
maximum of $L$ lies near $\pi_0$ and the correction is fairly minor; at
$\sigma_\pi /\pi_0 = 0.15$, for example, the distribution of true
parallaxes resembles a Gaussian with a peak at $\pi = 0.9 \pi_0$.
There is always a formal singularity near $\pi = 0$, but for small
relative error this occurs at distances so large as to be implausible
(see Fig. 1 of \citealt{lutzkelker})  \footnote{It is interesting to
note that the infinity near $\pi = 0$ is severe enough that the
cumulative distribution function is undefined unless the domain of the
probability distribution is restricted to be greater than some nonzero
parallax.}. However, as the relative error increases, the singularity
near $\pi = 0$ become dominant, and at $\sigma / \pi_0 = 0.25$, the
local maximum in $L$ disappears and there is no longer a unique estimate
of $\pi$ \citep{smith87a}.  In this framework, even a `4 $\sigma$'
parallax detection is likely to be a near-zero parallax which has been
bumped up to the observed value by observational error, because the
volume of space at small parallax is so large.  This difficulty at low
signal-to-noise might be called the `Lutz-Kelker catastrophe'.

Several of the stars studied here have parallaxes in the $\sim 3\sigma$
to $5 \sigma$ range, significant if the Lutz-Kelker bias is {\it not}
taken into account, but formally insignificant if the Lutz-Kelker
correction is taken at face value.  The formal insignificance does not
seem plausible, since there is other evidence for the relative proximity
of these stars, as follows.  (1) In some cases the CVs have
significantly larger proper motions than the reference stars.  (2) The
absolute magnitudes can often be constrained by other evidence, and in
any case these objects cannot have the very high luminosities implied by
great distances.  (3) The target stars, which have been selected {\it a
priori} as cataclysmic variables, generally have the largest parallaxes,
or close to it, among all the stars measured in the field.  The
Lutz-Kelker catastrophe arises because of random fluctuations operating
on a skewed underlying parallax population -- why should the target
stars be singled out?

We thus seek a way to incorporate our prior expectation that the objects
are {\it not} at extreme distance, in order to suppress the Lutz-Kelker
catastrophe.  \citet{lutzkelker} themselves remove the formal infinity
at zero parallax by arbitrarily imposing a lower limit to the parallax
(their $\epsilon$).  

Bayesian probability provides a formal structure for incorporating prior
information in parameter estimation \citep{loredo92} \footnote{A useful
expanded version of this article is available at
astrosun.tn.cornell.edu/staff/loredo/bayes/tjl.html}.  I therefore used
Bayesian inference to construct an {\it a posteriori} probability
density for the parallax, using as prior information the proper motion
together with an assumed velocity distribution, and the apparent
magnitudes together with a broad range of assumed absolute magnitudes.
This approach is cogently detailed by \citet{smith87a} and
\citet{smith87b}.  The formalism used here is nearly identical.

In order to make use of the proper motions, a distribution of velocities
must be assumed.  With the simplifying assumption of an isotropic
velocity distribution, the $\gamma$-velocities (i.e., mean systemic
radial velocities) give a measure of  the transverse velocity
distribution.  \citet{vanparadijs96} tabulated observed $\gamma$
velocities from the literature and
found, for non-magnetic systems, an overall dispersion of 33 km
s$^{-1}$, without strong dependence on subtype.   One might expect a
subtype dependence -- \citet{kolbstehle} predict that long-period
systems should be relatively young, and hence have a low velocity
dispersion, and indeed \citet{north02} find an extremely low velocity
dispersion among four systems they studied.  Shorter-period systems may
be more ancient, and hence have higher dispersions.   In order to check
this possibility with a possibly more homogeneous sample, I collected
$\gamma$-velocities of systems with $P_{\rm orb} < 2$ hr from published
MDM observations \citep{tpst, ccahvz, uvvyv1504, dnshort, thor03}, most
of which were not available to \citet{vanparadijs96}, and from another
ten systems for which results are in preparation.  All 28 stars in this
sample were observed with similar spectral resolution and calibration
procedures, which should minimize excess scatter.  The LSR-corrected
velocities had $\bar v = -9$ and $\sigma_v = 28$ km s$^{-1}$ (excluding
RZ Leo, which had a very poorly-measured orbit).  Both this result and
van Paradijs et al.'s estimate should be upper limits to $\sigma_v$,
since the emission lines do not necessarily track either star closely.
The velocity dispersion of the shorter period systems appears thus to be
quite small, probably only just consistent with the predictions of
\citet{kolbstehle} (their Fig.~3).  

\citet{smith87b} develops a formalism for computing the {\it a priori}
likelihood of a proper motion as a function of distance, when the parent
population has a triaxial Gaussian velocity distribution aligned with
Galactic coordinates, with dispersions $\sigma_U$, $\sigma_V$, and
$\sigma_W$ in the three principal directions (his equations 25 and 26).
I modified this formalism as follows.  Guided by the velocity dispersion
results noted above, I assumed the bulk of the CV population to have the
kinematics of Galactic K0 giants as tabulated in \citet{mihalbinn} (p.
423), with $(\sigma_U,\sigma_V,\sigma_W) = (31, 21, 16)$ km s$^{-1}$.  I
added to this a high-velocity tail with a normalization of 0.05 times
the bulk population, with (100, 75, 50) km s$^{-1}$, similar to the
subdwarfs.  Finally, I added a lower-velocity core, with 0.2 times the
normalization, with (24, 13, 10) km s$^{-1}$, similar to F0 dwarfs.
This composite probability density was evaluated over a range of
hypothetical true parallaxes.  The proper motion was adjusted to the
local standard of rest at each parallax before the probability density
was evaluated.  The uncertainty in the proper motion determination
(eqn.~27 in \citealt{smith87b}) was ignored.

% The proper motions were incorporated as follows.  At each hypothetical
% parallax $\pi$, the proper motion components of the dynamical
% LSR $\mu_{\rm X(LSR)}$ and $\mu_{\rm Y(LSR)}$ 
% were calculated, and the assumed $\sigma_v$ of the population 
% was converted to an equivalent $\sigma_{\mu} = \pi \sigma_v / 4.74$.  
% The velocity distribution, assumed isotropic about the LSR, 
% was modeled as the sum of three Gaussian functions in each 
% axis, one with the expected $\sigma_{\mu}$, one with 
% half this $\sigma$ and half the normalization, and another with
% double this $\sigma$ and again half the normalization.  The probability
% density associated with a given $\pi$ for a measured $\mu_X$ and 
% $\mu_Y$ was therefore composed of three terms of the form
% $$\hbox{constant} \times 
% {\exp ((\mu_{\rm X} - \mu_{\rm X(LSR)})^2/(2 \sigma_\mu^2)) \over \sigma_\mu}
% \times
% {\exp ((\mu_{\rm Y} - \mu_{\rm Y(LSR)})^2/(2 \sigma_\mu^2)) \over \sigma_\mu},
% $$
% with the $\sigma$ values and normalization constants adjusted as
% above.

The apparent magnitudes of many of these systems can also be used to
constrain the distance.  \citet{warn87} (also \citealt{warn}) showed
that the absolute magnitudes of dwarf novae at maximum light, corrected
for inclination, are strongly correlated with $P_{\rm orb}$.  For
$P_{\rm orb} < 2$ hr, the maximum magnitude generally occurs in
superoutburst, which is about a magnitude brighter than normal outburst;
I assumed this to be the case.  The optical colors of dwarf novae in
outburst are fairly close to zero, so the distinction between $m_{\rm
pg}$ and $m_{\rm V}$ is ignored.  The orbital inclinations for most of
the program objects are uncertain, so the inclination correction is not
known.  To account for this and unexpected scatter in the relation, and 
to avoid `assuming what we are trying to prove', the
magnitudes were assumed to follow a very broad Gaussian.   Sometimes
other constraints on the distance were available (e.g., from
detections of secondary stars), and again relatively broad probability
distributions were assumed to avoid steering the estimate too much.  
The notes on individual stars detail the adopted absolute magnitude constraints.

The parallax, proper motion, and magnitude information was
combined as follows.  

Bayes' theorem states, in general terms 
$$P(H|DI) \propto P(D|HI) P(H|I),$$ 
where $P$ represents a probability, $H$ the hypothesis, $D$
the data (in this case, the observed parallax), and $I$ the prior
information about the problem (constraints derived from the proper
motion and magnitude, and assumptions such as the normal distribution of
errors).  In this case $H$ is a hypothesized true parallax, e.g. `VY Aqr
has a true parallax of 8.2 mas', and we are asking for the likelihood
that $H$ is correct given $D$ (the measured parallax and its estimated
uncertainty) and $I$ (the proper motion and the assumptions about the
space velocity, the apparent magnitude and the assumptions about the
plausible range of absolute magnitudes, and the assumed normal
distribution of the experimental error).  The true parallax can be any
positive number, so we run the computations for a range of parallaxes
from near 0 up to large values (i.e., we vary $H$), creating a
continuous probability density.  This continuous probability density for
$P(H|DI)$ is exactly what we want: the relative likelihood of each {\it
true} parallax, {\it given} all the information we have available,
including our measurement.

The first factor on the right is the probability of obtaining 
our measured parallax, for the given true parallax and the prior
information.  Once the true parallax has been fixed, the assumption
of a normal error distribution yields
$$P(D|HI) \propto e^{-(\pi - \pi_0)^2 / 2 \sigma_\pi^2}.$$
Since the true parallax is held fixed at an assumed value, the 
proper motion and magnitude constraints do not affect this
factor.  

The second factor is the {\it a priori} probability of a particular
parallax, given only the prior information.  This itself is composed
of several factors.  The proper motion probability density is
included here.  The magnitude constraint is somewhat more difficult,
because of bias.  \citet{smith87b} treats the case of a Gaussian
distribution of absolute magnitudes for the type in question,
and formulates a Malmquist-type adjustment to the 
most likely absolute magnitude which accounts for the tendency to pick out 
absolutely brighter (hence more distant) members of a population
with a non-zero luminosity dispersion.  The correction
replaces the mean absolute magnitude $M_0$ with
$M^* = M_0 - 1.84 \sigma_M^2$.  Alternatively, one can formulate
a density by simply multiplying the volume element 
($\propto 1/\pi^4$) by the appropriate Gaussian weighting function centered
on $M_0$.  Somewhat counter-intuitively,
these approaches give the same probability density.  Because
of the very broad Gaussians used to characterize the luminosity priors
of most of the cataclysmics,
these functions end up resembling pure $1/\pi^4$ distributions,
except that the singularity as $\pi \rightarrow 0$ is eliminated
by the Gaussian cutoff in absolute magnitude.

The calculation for each star proceeded as follows.  The 
estimated $\pi_{\rm abs}$ and its external error, the estimated
absolute magnitude and catalogued apparent magnitude (as appropriate
for the type of CV and outburst state) were tabulated; the values
used are given in Table 4 and commented on further in the notes below. 
A grid of `true' parallax values $\pi$ was constructed, from 0.1 to 30 
mas in 0.1 mas increments.   This upper limit was chosen to be safely
larger than any of the measured parallaxes.  At each 
parallax, the probability density $P(D|HI)$ was
computed, and a cumulative distribution function was formed
from these.  The points at which the cumulative distribution
equaled 0.50, 0.159, and 0.841 were taken as the best
estimate of the parallax and the positive and negative 
`1-sigma' error bars.  Fig. 1 illustrates this process for 
VY Aqr, and the last columns of Table 3 give the 
results.

\section{Results}

Table 4 summarizes the parallax measurements and the 
distances derived from them.  A discussion of individual
objects follows. 

{\it VY Aqr:} The parallax alone, 
$\pi_{\rm abs} = 11.2 \pm 1.4$ 
mas, gives a distance near 89 pc.  The relative
error is small enough that the Bayesian adjustments to this
are fairly minor.  VY Aqr is an SU UMa 
star with an orbital period
of 0.06309(4) d \citep{uvvyv1504}.
The orbital inclination
is unknown, but emission lines in quiescence are strongly
double-peaked, suggesting $i > 50$ degrees, and there
is no hint of an eclipse, suggesting $i < 75$ degrees, so I
adopt $i = 63 \pm 13$ degrees, which combined with 
the orbital period yields $M_V ({\rm max}) = 5.8 \pm 0.7$ using 
the \citet{warn87} relation.
\citet{pattvy93} studied the outbursts; they do not quote 
$V_{\rm max}$ for normal outbursts, but from their figures 
I estimate this to be 10.8, with an uncertainty of at least
a few tenths of a magnitude.  In the Bayesian
calculation a  generous $\sigma = \pm 3$ mag was used, effectively 
unweighting the magnitude constraint.  The proper motion 
probability density peaks near $\pi = 8$ mas, 
corroborating the parallax.  The Bayesian distance 
estimate, 97 (+15,$-$12) pc, is slightly larger
than $1/\pi_{\rm abs}$, mostly because of correction
for the bias.  Recently, \citet{mennickent02}
detected the secondary star in the infrared, and
estimated a distance of 100 $\pm 10$ pc, in 
excellent agreement with the parallax-based 
estimate.

% The VY Aqr discussion was revised 2003 March ... better 
% reference star selection, made little difference.

{\it SS Aur:} \citet{harrison00} find $\pi_{\rm rel} = 3.74 \pm 0.63$
mas from the HST FGS, and estimate $\pi_{\rm abs} = 5.22 \pm 0.64$ mas,
which after a Lutz-Kelker correction yields a most probable
parallax of 4.97 mas.   The parallax here, $\pi_{\rm abs} = 4.8(1.1)$ mas, 
is not as precise but agrees nicely on the face of it.  

The prior information for the distance estimate (excluding
the HST FGS parallax for independence) is 
as follows.  The secondary star in SS Aur is an M1V 
\citep{friend90,harrison00}, and the spectral energy distribution
suggests that most of the $K$-band luminosity is from the secondary.
A normal M1V has $M_K = +5.5$ \citep{beuermann99}, and \citet{harrison00}
measure $K = 12.66$; I take this as the best prior distance
estimate, and assign a $\pm 1$ mag standard deviation.
For purposes of testing the Warner relation (discussed later),
I adopt $V_{\rm max} = 10.3$ from the GCVS and take 
$i = 39 \pm 8$ deg from a dynamical study of the secondary
by \citet{friend90}.
The MDM proper motion is small, ($\mu_\alpha,\mu_\delta$) =
($+1.7, -20.8$) mas yr$^{-1}$, in 
fair agreement with the the Lick Northern Proper
Motion Survey (\citealt{npm}, hereafter NPM) which gives 
(+8.2, $-15.3$) mas yr$^{-1}$, with a statistical error $\sim 5$ mas yr$^{-1}$,
in a frame referred to external galaxies.
\citet{harrison00} find (+8.3, $-$3.4) mas/yr, which disagrees
significantly in $\mu_\delta$ with the Lick and MDM determinations.
Two of their four
reference stars are in my field, and for those two I find 
($-$0.5, $-$4.9) mas yr$^{-1}$ for their star `SS Aur 2', and 
(6.4, $-$7.2) mas yr$^{-1}$ for their `SS Aur 12'.  
They do not comment on the proper motions of their reference stars;
I therefore adopt the MDM proper motion.

The proper motion-based parallax probability 
density peaks near $\pi = 3$ mas,
and the secondary star absolute magnitude constraint peaks just below
$\pi = 4$ mas.  Because of the substantial relative error of 
the parallax, the $\pi^{-4}$ correction enters strongly, 
giving a final
estimate of 283 ($+90,-60$) pc. This is just consistent with the
more accurate HST distance, 200 $\pm$ 30 pc, but adds little
weight. 

% discussion and numbers for SS Aur revised 2003 March 3.

{\it Z Cam:} The orbital period of this prototypical Z Cam star is
6.98 hr \citep{kraftzcam, thorzcam}.  Eclipses are not observed, but
the light curve shows structure at the orbital period, indicating that
the inclination is not too low.  Adopting $i = 65 \pm 10$ degrees, the
$M_{\rm max}$ - $P_{\rm orb}$ relation \citep{warn87} gives 
approximately $M_V \sim +4.3 (+0.6,-0.9)$,
which combines with $V_{\rm max} = +10.4$ (estimated from
\citealt{oppenheimeraavso})  to yield a most likely
$m - M = +6.1$, or $d = 165$ pc.  \citet{szkodywade81} classify the
secondary star as K7 and, assuming it is somewhat larger than the
ZAMS at that type, place the star at 200 pc, or $m - M = +6.5$.
Based on these I conservatively adopt a prior estimate of $m - M = 6.2 \pm 1.5$.  
The MDM proper motion is ($-17.1, -16.0$) mas yr$^{-1}$, while the
NPM gives ($-7.8, -9.0$), in fair agreement; I again adopt the MDM
measurement because of its small formal error.  

The measurement is $\pi_{\rm abs} = 8.9 \pm 1.7$ mas, or 112 pc at
face value.  The magnitude
constraint peaks around 6 mas, and the proper motion probability 
density peaks at just less than 3 mas (and the smaller NPM proper 
motion would push it still farther away).  The Bayesian distance 
estimate is 160 (+65, $-$37) pc; the parallax contributes 
significantly to bring the distance closer than the small proper 
motion would suggest.

% data checked and discussion revised 03 March 03.

{\it YZ Cnc:} The parallax measurement is barely significant
at  $4.4 \pm 1.7$ mas.  The MDM proper motion, ($+23.9$, $-47.7$) mas yr$^{-1}$,
agrees well with NPM, ($+18.2$, $-48.8$) mas yr$^{-1}$.  
The proper motion probability peaks near $\pi = 8$ mas.  The
orbital period of this SU UMa-type star was determined to be 2.08 hr
by \citet{shafterhessman88}, who argue that the orbital inclination
is around 40 deg.  From this we estimate $M_V {\rm (max)} = 
4.6 \pm 3.0$, and for normal outbursts we take $V = 12.0$ from
\citet{patterson79}.  The prior probability density based on the
magnitudes peaks just above $\pi = 3$ mas.   The final Bayesian
distance estimate is 256 ($-$70,+220) pc.  The Bayesian probability 
density is double-peaked, evidently an artifact of the 
low-weight, high-velocity tail on the assumed velocity distribution,
which extends some likelihood into the region where
$\pi^{-4}$ increases rapidly as $\pi$ decreases.  Interestingly,
\citet{dhillon00} did not detect a secondary in the infrared and
from this deduced $d > 290$ pc if the secondary's spectral type
is as expected.  The present result is consistent with this limit.
However, if the higher-velocity component of the velocity distribution
is removed, the upper distance limit is sharply curtailed, and the
Bayesian estimate becomes 222 ($+50$,$-42$) pc.   In this case the 
rapidly falling probability densities of the proper motion and 
parallax combine to cut off the long-distance end.

While this paper was in the final stages of preparation,
I became aware of an HST FGS parallax of YZ Cnc: \citet{harrison03a},
find $\pi_{\rm abs} =  3.34 \pm 0.45$ mas, and derive a Lutz-Kelker
corrected value near 3.1 mas, for a distance near $320 \pm 40$ pc.
The present result is at least nicely consistent with this much more precise
value.
 
% YZ Cnc data checked and discussion revised 2003 March 04.

{\it GP Com:} The distance is well-constrained by $\pi_{\rm abs} =
14.8 \pm 1.3$ mas.  Because GP Com is an unusual double-degenerate
helium CV \citep{marsh99}, the absolute magnitude constraint was
relaxed to $M_V = 10 \pm 8$, unweighting it almost completely in the 
distance estimate.  The proper motion is large --
($-337, +48$) mas yr$^{-1}$ in the present study, and 
($-343, +32$) mas yr$^{-1}$ in NPM, so the transverse
velocity $v_T = 110$ km s$^{-1}$ at $d = 1/\pi_{\rm abs}$.  
This places GP Com on the outer limits of the core velocity
distribution, but comfortably within the higher-velocity component.
The low-velocity component of the proper motion distribution
still carries some weight at 100 km s$^{-1}$, and it pulls the
Bayesian distance estimate closer than 
$1/\pi_{\rm abs}$, almost perfectly canceling the Lutz-Kelker bias.
The final Bayesian distance estimate is 68($+7,-6$) pc.

% GP Com was revisited and revised 03 March 3.  One high-PM
% reference star was taken out, tightening up the PM distribution
% of the references stars a bit.

{\it EF Eri:} Due to the faintness of this star, which was 
in its low state for all the parallax observations, the fit residuals 
for single observations are relatively large (15 mas on average).
This and the relatively sparse reference frame yielded a 
fairly uncertain parallax, $\pi_{\rm abs} = 5.5 \pm 2.5$ mas, 
so the Bayesian priors have a substantial effect.
The MDM proper motion (which we adopt) is quite large, 
(+118.9, $-$44.5) mas yr$^{-1}$,
with a formal error of 0.8 mas yr$^{-1}$; it disagrees 
significantly with the still larger NPM motion, 
(+144.2, $-55.1$) mas yr$^{-1}$.
The best photometric constraint comes from 
\citet{beuermanneferi}, who studied the low-state spectrum and 
identified the white dwarf contribution.  For $M_{\rm wd}$ =  0.7 M$_{\odot}$,
they estimate $d = 110$ pc.  We therefore form the constraint
by adjusting $m$ and $M$ to yield $m - M = 5.2 \pm 1.0$.  At 
$d = 1/\pi_{\rm abs}$, $v_T = 95$ km s$^{-1}$, so the assumed
high-velocity component in the population is important, giving
a final distance estimate of 
$d = 163(+66,-50)$ pc.  Both the proper motion and magnitude
priors push the estimate to lower distances than the parallax, and again
the Bayesian median distance is slightly more nearby than $1/\pi_{\rm abs}$
despite the Lutz-Kelker bias.  As in the case of YZ Cnc, the
Bayesian result depends critically on including a high-velocity
population in the velocity prior; removing the small admixture of 
high-velocity stars yields a considerably lower distance, $113 (+19,-16)$ pc.
The formal uncertainty is relatively small in this instance  because the
best compromise value lies far from the 
peaks of both the parallax and proper motion probability 
densities, so the net probability drops away quickly on each side of
its maximum.  All told, the parallax does tend to push the distance
well out beyond what one would guess from the proper motion alone,
and a little farther than the white-dwarf atmospheres argument would
suggest.

% EF Eri checked and revised 2003 March 4.

{\it AH Her:} A parallax is barely detected, $\pi_{\rm abs} = 3.0 \pm 
1.5$ mas.  The MDM proper motion is very small, ($0.0, +9.3$) mas
yr$^{-1}$, in excellent agreement with NPM which gives ($-0.8, + 10.8$).
\citet{bruch87} thoroughly studied the distance constraint from
the secondary star and concluded that the most likely distance
was 350 to 500 pc, the spectral type of the secondary being the 
biggest contributor to the uncertainty.  The \citet{warn87}
relation gives a very similar distance, using 
$i = 46 \pm 3$ deg from \citet{horneahher86},
and $V_{\rm max} = 11.6$ from \citet{spogli}.   For 
the magnitude prior I therefore adopt $m - M = 8.1 \pm 1.0$, equivalent
to $416$ pc = $1/2.4$ mas.  
The proper motion probability density peaks at an even greater
distance, though its median is around 370 pc.  Because of the 
large relative parallax error, and the small proper motion, this
object is heavily influenced by the $\pi^{-4}$ effect, and the 
Bayesian distance estimate is $d = 660(+270,-200)$ pc. 

% AH Her revised and updated 2003 March 4.

{\it AM Her:} The parallax, $\pi_{\rm abs} = 13.0 \pm 1.1$ mas, 
is accurate enough to dominate the distance estimate.  The MDM
proper motion is very substantial at ($-39.9, +29.7$) mas yr$^{-1}$;
it agrees well with the USNO B1.0 \citep{mon03}, which lists ($-41, +28$).
\citet{youngschneider81} detected the M4+V secondary and 
estimated $d = 71 \pm 18$ pc for a radius typical of the 
main sequence; for a somewhat larger secondary they estimated
93 pc.  From this we adopt $m - M = 4.5 \pm 1.5$ for the 
magnitude prior, which gives a probability peak almost identical
to $\pi_{\rm abs}$.  The proper motion probability density peaks
near $\pi = 9$ mas, and the final distance estimate is 
$79 (+8,-7)$ pc, just slightly farther than $1/\pi_{\rm abs}$.
The parallax adds weight and precision to previous estimates,
but does not revise them significantly.

% AM Her revised and checked 2003 March 4.

{\it T Leo:} The parallax, $\pi_{\rm abs} = 10.2 \pm 1.2$, 
yields $d = 98$ pc, and  
is accurate enough to dominate the distance determination.
The large MDM proper motion, ($-86.3,-50.2$) mas yr$^{-1}$,
compares with ($-87.9, -66.1$) in the NPM, and gives a
probability density peaking around 70 pc.  
\citet{shafterszkody84} measured the 
84.7 min orbital period, and estimated $28 \le i \le 65$ deg, 
from which the \citet{warn87} relations yield 
$4.6 \le M_{V{(\rm max)}} \le 6.2$; to be conservative
I take $M_{V{(\rm max)}} = 5.4 \pm 3.0$.  At supermaximum
this SU UMa star reaches around $V = 10.0$ \citep{howelltleo,katotleo}, 
so I adopt $V = 11$ for
ordinary maxima, yielding a most probable distance of 132 pc
from the magnitude constraint alone.  The final Bayesian
distance estimate is 101 ($+13,-11$) pc.   \citet{dhillon00}
use their non-detection of the secondary star in the 
infrared to establish $d > 120$ pc; the parallax argues
for a distance near their lower limit, and suggests that the
secondary lurks just below their sensitivity.

% T Leo revised and checked 2003 March 4.

{\it GW Lib:} GW Lib's orbital period,  $P_{\rm orb} = 76.8$ min, 
is the shortest known among dwarf novae with normal-composition 
secondaries \citep{dnshort}.  The parallax, $\pi_{\rm abs} = 11.5 \pm 2.4$,
is not particularly accurate.  The MDM proper motion,
($-62,+28$) mas yr$^{-1}$ is quite substantial,
and agrees fairly well with ($-58,+20$) listed in   
USNO B1.0.  This dwarf nova has only been seen
in outburst once, and it was poorly observed;
\citet{dnshort} estimated $d = 125$ pc from the 
available outburst information.  The white dwarf
is visible in the spectrum.  \citet{szkodygw} fit
$\log g = 8$ models of white dwarf atmosphere to HST 
ultraviolet data, and found distances of 171 and 
148 pc depending on the temperatures used,  
but did not constrain $\log g$ independently and
did not explore the distance parameter space.  For the
present study the absolute magnitude prior was 
set arbitrarily to $m - M = +5.5 \pm 3$ to match the 
(poorly determined) outburst absolute magnitude,
effectively unweighting this constraint.
The sizable relative parallax error creates a 
substantial Bayes-Lutz correction, but the large
proper motion counteracts this, leading to a 
final distance estimate of $104 (+30, -20)$ pc.

{\it V893 Sco:} This relatively bright dwarf nova
had been lost until its identification was clarified by
\citet{katov893i}.
This field is sparsely observed, yielding 
$\pi_{\rm abs} = 7.4 \pm 2.4$ mas.  The MDM proper motion is
($-53,-53$) mas yr$^{-1}$.
The period is 1.823 hr, and the inclination is constrained by eclipses 
\citep{bruchv893}, leading to an estimated $M_V$ at maximum of 
$6.0 \pm 2.0$ (we adopt 2 mag rather than 3 mag for the 
uncertainty because of the quality of the inclination constraint), 
and \citet{katov893ii} find that normal 
outbursts reach $V = 12.5$; the probability density from
this constraint peaks around $\pi = 5$ mas.  The 
proper motion probability density peaks around 
11 mas.  The final distance estimate is
$155 (+58, -34)$ pc.  

{\it WZ Sge:} This is the best-determined parallax in
this study, $\pi_{\rm abs} = 23.2 \pm 0.8$ mas.

The parallax is accurate enough that the Bayesian
priors have almost no effect, but the absolute magnitude
estimate will be used later, so details are given here.
All of WZ Sge's outbursts are superoutbursts, and they
reach $V = 8.2$ \citep{pattwz01ob}, which would
imply a normal outburst magnitude of $V = 9.2$, if
normal outbursts actually occurred.  \citet{spruit}
find $i = 77 \pm 2$ degrees, and I double this uncertainty
to be conservative.  The inclination-adjusted $M_V$-$P_{\rm orb}$ 
relation then implies $M_V = 6.7$.  For the Bayesian
estimate an uncertainty of 3 mag was assumed, effectively
unweighting this constraint.  The MDM proper motion is 
($+74.3,-19.5$) mas yr$^{-1}$, leading to a proper 
motion probability density peaking at $\pi = 15$ mas.  
The final distance estimate is $43.3 (+1.6, -1.5)$ pc.  

The literature contains many other estimates for the 
distance of WZ Sge.  \citet{smak93} estimated
$48 \pm 10$ pc from the flux of the white dwarf, for
which he adopts $M_{\rm wd} = 0.45 M_{\odot}$. 
\citet{sionwz} model an HST ultraviolet spectrum
with a $\log g = 8$ white dwarf, and find 
69 pc for the distance.  However, \citet{spruit}
point out that the temperature and gravity are
highly degenerate in this kind of fit; largely
from a study of the stream dynamics they adopt
$\log g = 9$ for the white dwarf, and argue that the
\citet{sionwz} distance should be adjusted downward
to 48 pc.  The short distance determined here supports
their interpretation, and suggests that the white
dwarf in WZ Sge is relatively high-gravity (hence
massive).   WZ Sge is evidently the closest known 
cataclysmic binary.

While this paper was in the final stages of 
preparation, two other measurements of the parallax
of WZ Sge came to my attention.  First, C. Dahn
(private communication) kindly passed along a USNO
$\pi_{\rm rel}$ for WZ Sge closely agreeing with the
present determination.  Second, \citet{harrison03a} announced
an even more precise $\pi_{\rm abs} = 22.97 \pm 0.15$ mas
from the the HST Fine Guidance System, which also agrees
accurately with this one.  With three independent 
determinations giving essentially the same value, 
we can be very confident about the distance to 
WZ Sge.

% check Smak and Spruit references.
% Smak, J. 1993, Acta Astronomica, 43, 101
% Spruit, H. C., and Rutten, R. G. M., 1998, MNRAS, 299,
%  768   -- also around 48 pc from WD fitting.

% WZ discussion written 2003 March 4.

{\it SU UMa:} 
The proper is small, $(+2,-16)$ mas yr$^{-1}$, which compares with 
$(+11.2,-25.2)$ mas yr$^{-1}$ in the Lick NPM.
The photometric constraint is based on 
$V = +12.0$ for normal outbursts \citep{rosensu},
and the Warner relation with 
an ill-constrained binary inclination of 60 degrees,
based on the modest velocity amplitude \citep{thorwade86}.
The parallax, $\pi_{\rm abs} = 7.4$ mas,
is not well-determined, and the distance estimate 
depends critically on the error estimate adopted;
the scatter about the best fit indicates $\sigma = 1.4$
mas, while the scatter of six comparison stars within
2 mag of SU UMa gives 2.4 mas.  Adopting the smaller
error yields a Bayesian estimate of 189 $(+79,-43)$ pc;
if the larger error is adopted, the Lutz-Kelker 
correction increases the estimate to 403 $(+230,-162)$ pc.
While the distance is disappointingly indeterminate,
120 pc is a reasonable lower limit.

{\it HV Vir:} This star resembles WZ Sge in that it
outbursts only rarely and with large amplitude.  
Because of its faintness, the parallax
is somewhat uncertain at $\pi_{\rm abs} = 5.8 \pm 2.2$ mas.
The proper motion is modest, $(+19,-12)$ mas yr$^{-1}$.
in good agreement with $(+22, -8)$ in USNO B1.0.  
The superoutburst light curve \citep{leib94} reaches
$V = 11.5$, indicating that if normal outbursts occurred
(which they apparently don't), they would reach $\sim 12.5$.  
The orbital period, from low-state photometry, is 
0.05799 d, and the presence of a low-state modulation 
suggests a substantial orbital inclination, yielding
an estimated
$M_V = 5.9 \pm 3$.  The resulting photometric constraint
peaks near $\pi = 5$ mas.  The small proper motion
puts the star more distant, with a peak near $\pi = 3$ 
mas.  The rather weak parallax determination leads to 
a substantial $\pi^{-4}$ effect, and a final 
Bayesian distance estimate of 460 ($+530, -180$) pc.
For comparison, \citet{szkodyhv} estimate a distance in the 
400 -- 550 pc range from the white dwarf's UV
continuum.  WZ Sge is about 4.5 mag brighter than
HV Vir; assuming they are identical yields a distance
estimate near 350 pc, in reasonable agreement with the 
Bayesian estimate.

\section{Discussion}

Although the distance scale for cataclysmics has been
uncertain (as noted in the Introduction),
over the years some `conventional wisdom' has
grown up around cataclysmic distances, based on detections of
secondary stars, kinematical evidence, and the like.  
How well does the conventional wisdom bear up?

{\it Dwarf Novae.} We can use the dwarf novae for which we 
have usable distance estimates to test the 
$M_V({\rm max}$-$P_{\rm orb}$ correlation
\citep{warn87,warn}.  
Table 5 and Fig.~3 show this test\footnote{In order to avoid circular
arguments the distance estimates employed in this comparison
are based on the parallax and proper motion
only, without the magnitude constraint.}.  
There are several complications,
as follows: (1) Warner's correlation depends on a correction for orbital 
inclination, which is often poorly known; the text of the 
previous section gives the evidence used to constrain
$i$.  The values of $M_V$ (pred) in Table 5 and Fig.~3 are for
the assumed inclination, rather than 
corrected to a particular fiducial inclination, so they should
be directly comparable to observation.
The error bars on the predicted $M_V$ reflect only the
uncertainty in the inclination, and {\it ignore} any `cosmic scatter'
in the relation. (2) The superoutbursts of SU UMa stars are about 
one magnitude brighter than the normal outbursts.  For those SU UMa stars
for which I was unable to find literature references to $V_{\rm max}$ for 
normal outbursts I've added 1 mag to the superoutburst maximum.
In the WZ Sge, HV Vir, and GW Lib,
normal outbursts are not observed, but the same correction was adopted.  

In view of the crude assumptions -- especially the arbitrary
1-magnitude correction between normal and superoutbursts, and the
unreliability of some of the inclinations -- the agreement appears to be
satisfactory.  Although the data appear too sparse to uncover
the relationship independently, none of the stars are markedly
discrepant.  The rarely-outbursting, large-amplitude objects -- WZ Sge,
HV Vir, and GW Lib -- do not disagree dramatically with expectations.
Even so, WZ Sge is measured to be somewhat farther away than one would
predict on the basis of the relation, which is puzzling in that it has
an accurate magnitude, a well-constrained inclination, and a very
accurately-determined distance.  Adopting the actual superoutburst
$V_{\rm max}$ for the maximum magnitude would make the discrepancy
worse.  It is possible that the disk in WZ Sge expands to be unusually
large during its outbursts, increasing its intrinsic brightness beyond
expectation, or that the expression used by Warner for the inclination
correction becomes inaccurate at high inclinations.  SS Aur is also
slightly discrepant (less than two standard deviations), in the sense
that the predicted absolute magnitude is fainter than the empirical one.
The inclination is not strongly constrained, but is already assumed to
be fairly modest, and even adopting a face-on inclination would not
brighten the predicted magnitude appreciably.  However, the more
accurate HST FGS parallax \citep{harrison00} brings the empirical
absolute magnitude to 3.8, much closer to the predicted value.  

\citet{harrison03a} discuss the $M_V$ (max) - $P_{\rm orb}$ relation
at greater length using the HST parallaxes, and confirm that
the relationship appears to hold.

{\it AM Her stars.}  The parallax of AM Her agrees well
with distances based on the spectrum of the secondary.  EF Eri
is not accurately determined but comes in a little farther away than
the white dwarf atmosphere \citep{beuermanneferi} would suggest.  
\citet{harrison03}
have recently measured and modeled infrared spectra and light curves 
of EF Eri, but do not comment on how the models are normalized to the
data (that is, the distance); the infrared light curve is quite
complicated and so model dependencies are likely to creep into 
such determinations in any case.

{\it Helium CVs.} GP Com was the only helium CV included,
but it appears to be the first to have an accurate distance determination.
Taking our measured $V = 16.1$ as typical, the measured
distance modulus $m - M = 4.2 \pm 0.2$ yields $M_V = +11.9$.

\section{Conclusions} 

The main conclusions are as follows.

(1) As USNO92 assert, interestingly accurate parallaxes can be
derived without special equipment, provided the instrumentation
is stable.

(2) Over the years a fair amount of conventional wisdom has
grown up around cataclysmic distances, based on detections of
secondary stars, kinematical evidence, and the like.  This study
largely corroborates this conventional wisdom; a one-line
summary might be `no big surprises'.  

(3) Even so, there are some small surprises.  WZ Sge is a little
closer than some previous estimates had suggested, and a little
farther away than predicted by the $M_V$(max) - $P_{\rm orb}$
relation.   Although the result for EF Eri is imprecise, 
it appears to be a little farther away than anticipated. 

(4) GP Com is evidently the first helium CV with a reliable distance. 
It is intrinsically faint ($M_V = +11.9$).  Furthermore, its
transverse velocity is 110 km s$^{-1}$, outside the rather
small velocity dispersion of the main CV population.

{\it Acknowledgments.} Special thanks go to Dave Monet for
encouragement and free advice, which was infinitely many times more
valuable than its price, and for his role in pioneering this powerful 
technique.  Conard Dahn communicated a USNO parallax for WZ Sge
while this paper was in preparation, and it was a great 
confidence-builder to find it was essentially identical to that 
presented here.  Also, Tom Harrison kindly 
communicated the wonderfully accurate HST parallaxes just as I was 
completing this paper.  Tom Marsh suggested GP Com as a target.
After I had begun development of the Bayesian distance
estimation techniques I discovered that Haywood Smith had already
explored this approach thoroughly, and he offered much thoughtful
advice.  Graduate students Cindy Taylor and Bill Fenton took 
data on several observing runs.  The MDM staff and director
put up with extra instrument changes so that this program could be 
shoehorned into time otherwise 
used for  spectroscopy.   Joe Patterson made some thoughtful
comments.  Last but not least, I thank the NSF 
(AST 9987334 and AST 0307413) for support.

\clearpage

\begin{deluxetable}{lrrrl}
\tabletypesize{\scriptsize}
\tablewidth{0pt}
\tablecolumns{5}
\tablecaption{Journal of Observations}
\tablehead{
\colhead{Star} & 
\colhead{$N_{\rm ref}$} & 
\colhead{$N_{\rm meas}$}  & 
\colhead{$N_{\rm pix}$} &
\colhead{Epochs} \\ 
}
\startdata
  VY Aqr  &   15  &  22  & 139  & 1997.71(17) , 1997.95(5) , 1998.44(8) , 1998.69(11) , 1999.43(21) , \\
           &       &      &      &  1999.79(27) , 2000.50(24) , 2002.81(8) , 2003.46(18) \\
  SS Aur  &   59  & 148  & 132  & 1997.71(9) , 1997.96(15) , 1999.05(44) , 1999.79(37) , 2000.03(15) , \\
           &       &      &      &  2000.26(6) , 2002.05(6) \\
   Z Cam  &   11  &  32  &  59  & 1997.95(19) , 1998.21(6) , 1999.05(19) , 1999.79(11) , 2000.03(4) \\
  YZ Cnc  &   25  &  50  &  83  & 1997.95(5) , 1998.21(6) , 1999.05(18) , 1999.79(5) , 2000.03(9) , \\
           &       &      &      &  2000.26(30) , 2002.05(10) \\
  GP Com  &   13  &  31  & 152  & 1999.43(19) , 2000.03(24) , 2000.26(23) , 2000.50(12) , 2001.24(20) , \\
           &       &      &      &  2001.39(24) , 2003.08(17) , 2003.46(13) \\
  EF Eri  &   13  &  28  & 105  & 1997.71(19) , 1997.95(7) , 1999.05(14) , 1999.79(9) , 2000.03(26) , \\
           &       &      &      &  2002.05(11) , 2003.08(19) \\
  AH Her  &   17  &  80  &  93  & 1998.21(17) , 1998.44(14) , 1998.69(3) , 1999.43(30) , 2000.26(7) , \\
           &       &      &      &  2001.24(6) , 2001.39(16) \\
  AM Her  &   24  &  54  & 105  & 1997.71(11) , 1998.21(19) , 1998.44(14) , 1998.69(5) , 1999.43(25) , \\
           &       &      &      &  2000.26(19) , 2000.50(9) , 2001.39(3) \\
   T Leo  &   15  &  24  & 130  & 1997.95(1) , 1998.21(12) , 1998.44(6) , 1999.05(7) , 1999.43(11) , \\
           &       &      &      &  2000.03(41) , 2000.26(18) , 2001.24(4) , 2002.05(13) , 2003.08(17) \\
  GW Lib  &   41  & 110  &  70  & 2000.26(19) , 2000.50(15) , 2001.24(9) , 2001.39(19) , 2001.46(2) , \\
           &       &      &      &  2003.08(6) \\
V893 Sco  &   47  & 187  &  83  & 2000.26(22) , 2000.50(21) , 2001.24(10) , 2001.39(13) , 2003.46(17) \\
  WZ Sge  &   45  &  77  & 162  & 1997.71(12) , 1998.44(15) , 1999.43(44) , 1999.79(15) , 2000.50(36) , \\
           &       &      &      &  2001.39(21) , 2002.81(19) \\
  SU UMa  &   10  &  25  &  90  & 1997.96(17) , 1998.21(2) , 1999.05(14) , 1999.79(2) , 2000.03(30) , \\
           &       &      &      &  2000.26(9) , 2002.05(9) , 2003.08(7) \\
  HV Vir  &   12  &  32  & 111  & 1998.21(13) , 1998.44(10) , 1999.05(29) , 1999.43(29) , 2000.26(20) , \\
           &       &      &      &  2001.24(4) , 2001.39(6) \\
  LHS429  &   17  &  41  &  58  & 1998.21(4), 1998.44(8), 1999.43(15), 2000.26(6), 2000.50(5), \\
          &       &      &      & 2001.39(20) \\
  LHS483  &   40  &  84  &  62  & 1997.71(6), 1998.44(5), 1999.43(7), 1999.79(3), 2000.50(11), \\
          &       &      &      &  2001.39(20), 2002.81(10) \\
LHS1801+2  &   25  &  59  &  65  & 1997.71(14), 1998.21(6), 1999.04(21), 1999.79(9), 2000.03(5), \\
           &       &      &      & 2002.05(10) \\
 LHS1889  &   46  &  68  &  63  & 1999.05(29), 1999.79(8), 2000.03(8), 2001.24(10), 2002.81(8) \\
 LHS3974  &   22  &  51  &  63  & 1997.71(14), 1997.96(16), 1998.68(5), 1999.79(22), 2002.81(6) \\
\enddata
% \tablenotetext{a}{Number of spectra.}
\tablecomments{Overview of the data included in the parallax solutions. 
$N_{\rm ref}$ is the number
of reference stars used to define the plate solution, $N_{\rm meas}$ is the total
number of stars measured, and  $N_{\rm pix}$ is the number of images used.  The epochs
represent different observing runs, and the numbers in parentheses are the number of
images included from each run.}
\end{deluxetable}

\clearpage
\begin{deluxetable}{rrrrrrrrrr}
\tabletypesize{\scriptsize}
\tablewidth{0pt}
\tablecolumns{10}
\tablecaption{Positions, Magnitudes, Parallaxes, and Proper Motions}
\tablehead{
\colhead{$\alpha$} &
\colhead{$\delta$} &
\colhead{Weight} &
\colhead{$\sigma$} &
\colhead{$V$} &
\colhead{$V-I$} &
\colhead{$\pi_{\rm rel}$} &
\colhead{$\mu_X$} &
\colhead{$\mu_Y$} &
\colhead{$\sigma_{\mu}$} \\
\colhead{[J2000]} &
\colhead{[J2000]} &
\colhead{} &
\colhead{[mas]} &
\colhead{} &
\colhead{} &
\colhead{[mas]} &
\colhead{[mas y$^{-1}$]} &
\colhead{[mas y$^{-1}$]} &
\colhead{[mas y$^{-1}$]} \\ 
}
\startdata
\cutinhead{VY Aqr:}
 21:12:05.70 &  -8:47:48.5 & 0 &  17 & 17.97 &  0.90 & $-0.3 \pm 1.9$ & $  -6.6$ & $  -2.8$ & 0.8\\ 
 21:12:08.65 &  -8:47:47.0 & 1 &   8 & 15.18 &  0.97 & $ 1.1 \pm 0.8$ & $   0.2$ & $   7.6$ & 0.4\\ 
 21:12:09.90 &  -8:47:53.1 & 1 &   8 & 17.27 &  1.05 & $ 1.3 \pm 0.9$ & $ -10.3$ & $   0.9$ & 0.4\\ 
 21:11:56.61 &  -8:48:05.7 & 1 &   9 & 16.36 &  1.00 & $-1.5 \pm 1.0$ & $   5.4$ & $  -4.4$ & 0.4\\ 
 21:12:17.21 &  -8:47:56.6 & 1 &   9 & 17.32 &  1.28 & $-0.6 \pm 0.9$ & $   6.4$ & $  -1.9$ & 0.4\\ 
 21:12:13.26 &  -8:48:14.6 & 1 &  10 & 16.98 &  1.04 & $-1.3 \pm 1.1$ & $   0.3$ & $ -17.3$ & 0.5\\ 
\enddata
\tablecomments{Parameters for all measured stars in all the fields.
Program stars are marked with an asterisk.  The fourth star listed
in the LHS 1889 field proved to be a hitherto unknown L3.5 dwarf.
The celestial coordinates are from mean CCD images and are referred
to the USNO A2.0, which is in turn aligned with the ICRS; the epochs
of the images used are typically around 1998.  Coordinates should
be accurate to $\sim 0''.3$ external and somewhat better than this
internally.   A 1 or 0 in the next
column indicates whether a star was used as a reference star.  The
next column gives the scatter around the best astrometric fit (see
text); in a few cases these are very large (e.g. close pairs which
were intermittently resolved).  The $V$ and $V-I$ colors come next,
with typical external uncertainties of 0.05 mag and internal 
consistency somewhat better than that.  Next come the fitted 
parallaxes, proper motions in $X$ and $Y$, and the uncertainty in the
proper motion (per coordinate). Full table available in the 
electronic version of this paper.}

\end{deluxetable}

\begin{deluxetable}{llrrrrr}
\tabletypesize{\small}
\tablewidth{0pt}
\tablecolumns{7}
\tablecaption{LHS Stars Re-Observed}
\tablehead{
\colhead{Star} & 
\colhead{Program} & 
\colhead{$\pi_{\rm rel}$} & 
\colhead{$\mu_{X{\rm (rel)}}$}  & 
\colhead{$\mu_{Y{\rm (rel)}}$}  & 
\colhead{$V$} &
\colhead{$V - I$} \\ 
\colhead{} & 
\colhead{} & 
\colhead{(mas)} & 
\colhead{(mas yr$^{-1}$)}  & 
\colhead{(mas yr$^{-1}$)}  & 
\colhead{} &
\colhead{} \\ 
}
\startdata
LHS429 & USNO & 153.9(0.7) & $-$814.7(0.6) & $-$869.1(0.6) & 16.85 & 4.54 \\
       & MDM  & 154.5(2.4) & $-$811.5(1.0) & $-$865.5(0.9) & 16.80 & 4.63 \\
\\
LHS483 & USNO & 56.7(0.8) & 1066.1(0.4) & $-$78.5(0.4) & 17.00 & 1.15 \\
       &  MDM & 58.8(1.2) & 1059.6(0.8) & $-$78.9(0.8) & 17.15 & 1.17 \\
\\
LHS1801 & USNO & 31.8(1.5) & 13.9(1.0) & $-$532.7(1.0) & 17.21 & 0.93 \\
       & MDM  & 30.6(1.6) & 11.9(0.9) & $-$528.1(0.9) & 17.20 & 0.95 \\
\\
LHS1802 & USNO & 36.4(1.1) & 12.6(1.1) & $-$528.2(1.1) &  15.46 & 2.86 \\
       & MDM  & 30.8(1.3) & 11.4(0.7) & $-$524.8(0.7) & 15.45 & 2.91 \\
\\
LHS1889 & USNO & 52.8(0.9) & 358.7(0.6) & $-$578.4(0.6) & 16.56 & 0.99 \\
        & MDM &  49.6(1.5) & 350.9(0.8) & $-$590.2(0.8) & 16.57 & 1.00 \\
\\
LHS3974 & USNO & 13.2(0.7) & 540.9(0.4) & 19.3(0.4) & 17.43 & 2.74 \\
        & MDM  & 11.8(1.3) & 539.6(0.6) & 17.0(0.6) & 17.42 & 2.81 \\
\enddata
\end{deluxetable}

\clearpage

\begin{deluxetable}{lrrrr}
% \tabletypesize{\scriptsize}
\tabletypesize{\small}
\tablewidth{0pt}
\tablecolumns{5}
\tablecaption{Parallaxes, Proper Motions, and Distances}
\tablehead{
\colhead{Star} & 
\colhead{$\pi_{\rm rel}$} & 
\colhead{$\pi_{\rm abs}$} & 
\colhead{$[\mu_\alpha, \mu_\delta]_{\rm rel}$} &
\colhead{$1/\pi_{\rm abs}$} \\ 
\colhead{} &
\colhead{$d_{\rm LK}$} &
% \colhead{$d(\pi)$} &
\colhead{$d(\pi,\mu)$} &
\colhead{$(m - M)$ prior} & 
\colhead{$d(\pi, \mu, m-M)$} \\
}
%\colhead{} &
%\colhead{(mas)} &
%\colhead{(mas)} &
%\colhead{(mas yr$^{-1}$)} &
%\colhead{(pc)} &
%\colhead{(pc)} &
%\colhead{(pc)} &
%\colhead{(mag)} &
%\colhead{(pc)} \\
%}
\startdata
  VY Aqr & $10.2 \pm 1.4 [1.1]$  & 11.2(1.4) & $+33.9,-38.4 (0.8)$ & $89(+13,-10)$ \\
 & 96 & $97(+15,-12)$ & 10.8; 5.8; 3.0  & $97(+15,-12)$ \\
\\
  SS Aur & $4.0 \pm 0.8 [1.2]$  & 4.8(1.1)  & $+1.7,-20.8 (0.6)$  & $208(+62,-39)$ \\
 & 300 & $300(+130,-75)$ & 12.66; 5.5; 1.0 & $283(+90,-60)$ \\
\\
   Z Cam & $7.7 \pm 1.8 [1.3]$  & 8.9(1.7)  & $-17.1,-16.0 (2.0)$ & $112(+27,-18)$ \\
 & 137 & $164(+88,-42)$ & 10.4; 4.3; 1.5 & $163(+68, -38)$ \\
\\
  YZ Cnc & $3.4 \pm 1.5 [1.7]$  & 4.4(1.7)  & $+23.9,-47.7 (0.9)$ & $230(+140,-60)$ \\
 & \nodata & $260(+320,-70)$ & 12.0; 4.6; 3.0 &$256(+290,-70)$ \\
\\
  GP Com & $13.5 \pm 0.7 [1.3]$ & 14.8(1.3) & $-336.8,+47.7 (0.5)$ & $69(+7,-6)$   \\
 & 70 & $68(+7,-6)$ & 15.7;10.0;8.0 & $68(+7,-6)$ \\
\\
  EF Eri & $3.9 \pm 1.6 [2.5]$  & 5.5(2.5) & $+119,-45 (0.8)$ & $182(+152,-57)$ \\ 
 & \nodata & $229(+85,-83)$ & [5.2];1.0 & $163(+66,-50)$  \\
\\
  AH Her & $1.7 \pm 1.0 [1.5]$  & 3(1.5)  & $+0.0, +9.3 (0.7)$ & $330(+330,-110)$  \\
 & \nodata & $1800(+2100,-900)$ & [8.1]; 1.0& $660 (+270,-200)$ \\
\\
  AM Her & $12.0 \pm 1.0 [1.1]$ & 13.0(1.1) & $-39.9, +29.7 (0.9)$ & $78(+7,-6)$    \\
 & 79 & $79(+8,-6)$ & [4.5]; 1.5  & $79(+8,-6)$ \\
\\
   T Leo & $9.1 \pm 0.7 [1.2]$  & 10.2(1.2)  & $-86.3, -50.2 (0.4)$& $98(+13,-10)$  \\
 & 104 & $101 (+13.,-11)$ & 11.;5.4;3. & $101(+13,-11)$ \\
\\
  GW Lib & $10.4 \pm 2.4 [1.8]$  & 11.5(2.4) & $-61.5, +27.8 (1.2)$& $87(+23, -15)$  \\
 & 112 & $104(+30,-20)$ & 9.0;3.5;3.0 & $104(+30,-20)$ \\
\\
V893 Sco & $5.9 \pm 2.4 [2.2]$  & 7.4(2.4)  & $-52.6, -52.8 (1.2)$& $135(+65,-33)$  \\
 & \nodata & $153(+68,-35)$ & 12.5;6.0;2.0  & $155(+58,-34)$  \\
\\
 WZ Sge & $22.0 \pm 0.6 [0.8]$ & 23.2(0.8) & $+74.3, -19.5 (0.5)$& $43.1(+1.5,-1.4)$    \\
 & 43.3 & $43.3(+1.6, -1.5)$ & 9.2;6.7;3.0  & $43.3(+1.6,-1.5)$  \\
\\
 SU UMa & $6.6 \pm 1.4 [1.7]$ & 7.4(1.7) & $+1.7, -16.1 (1.5)$& $135(+40,-25)$  \\
 & 194 & $269(+240,-99)$ & 12.0;5.4;3.0  & $260(+190,-90)$  \\
\\
 HV Vir & $4.9 \pm 2.2 [2.1]$  & 5.8(2.2)  & $+19, -12 (1.9)$& $172(+105,-47)$ \\
 & \nodata & $600(+1100,-280)$ & $12.5;5.9;3.$ & $460 (+530,-180)$  \\
\enddata
\clearpage
\tablecomments{Parallaxes (in mas), proper motions (in mas yr$^{-1}$), 
and distance estimates (in pc).  
Two lines are given for each star.  In the first line, 
the uncertainties given for the relative parallaxes are derived from the 
goodness of fit, and the square-bracketed quantities are the uncertainties
derived from the scatter of the reference stars (see text).  
The uncertainties in the proper motion do not take the uncertainty in the 
zero point into account, which is of order 5 mas yr$^{-1}$ in most
cases (see text).
The last column gives the distance based on the absolute parallax alone
with no corrections.
The second line for each star lists the following: 
{\it Col 2.} The most probable value of the distance
based on the absolute parallax and a Lutz-Kelker
correction (i.e., Bayesian with minimal priors); no confidence interval
is given because the cumulative distribution function is not normalizable
in this case. {\it Col. 3}  The 
Bayesian estimate considering the parallax and  proper motion prior only.
{\it Col. 4} The distance priors, expressed in magnitudes.  Where three
numbers are given they are the apparent magnitude, assumed absolute magnitude,
and 1-sigma combined uncertainty (generally taken to be quite large to 
avoid undue circularity); where two numbers are given, they signify an assumed
distance modulus $m - M$ and its associated uncertainty.  In these cases,
the distance prior is not associated with a literal apparent and absolute
magnitude.  {\it Col. 5} The Bayesian estimator taking into account 
parallax, proper motion, and the prior information from the previous
column.  In all these estimates, the errors quoted reflect the 16- and
84-percentile points in the cumulative distribution function.}
\end{deluxetable}

\clearpage

\begin{deluxetable}{lrrrrr}
\tabletypesize{\small}
\tablewidth{0pt}
\tablecolumns{6}
\tablecaption{Measured and Theoretical Absolute Magnitudes}
\tablehead{
\colhead{Star} & 
\colhead{$P_{\rm orb}$} &
\colhead{Inclination} &
\colhead{$V_{\rm max}$} & 
\colhead{$M_V$(max) (pred)} &
\colhead{$M_V$(max) (meas)} \\
\colhead{} &
\colhead{(hr)} &
\colhead{(deg)} &
\colhead{} &
\colhead{} &
\colhead{} \\
}
\startdata
GW Lib    &  1.28 & $11 \pm 10$ & 10.0 & $ 4.9(+0.5,-0.6)$ & $ 4.4(+0.1,-0.0)$ \\ 
WZ Sge    &  1.36 & $77 \pm  5$ &  9.2 & $ 6.0(+0.1,-0.1)$ & $ 6.7(+0.6,-0.4)$ \\ 
HV Vir    &  1.39 & $60 \pm 10$ & 12.5 & $ 3.6(+1.4,-2.3)$ & $ 5.5(+0.6,-0.4)$ \\ 
T Leo     &  1.42 & $47 \pm 19$ & 11.0 & $ 6.0(+0.3,-0.3)$ & $ 5.0(+0.8,-0.4)$ \\ 
VY Aqr    &  1.51 & $63 \pm 13$ & 10.8 & $ 5.9(+0.3,-0.3)$ & $ 5.6(+0.9,-0.5)$ \\
V893 Sco  &  1.82 & $71 \pm  5$ & 12.5 & $ 6.6(+0.6,-0.8)$ & $ 6.1(+0.4,-0.3)$ \\ 
YZ Cnc    &  2.08 & $40 \pm 10$ & 12.0 & $ 4.9(+0.7,-1.7)$ & $ 4.7(+0.3,-0.2)$ \\ 
SS Aur    &  4.39 & $39 \pm  8$ & 10.3 & $ 2.9(+0.6,-0.8)$ & $ 4.0(+0.2,-0.2)$ \\
Z Cam     &  6.98 & $65 \pm 10$ & 10.4 & $ 4.3(+0.6,-0.9)$ & $ 4.3(+0.7,-0.5)$ \\
\enddata
\tablecomments{The sources for the orbital inclinations and apparent $V$ magnitude
in {\it normal} outburst are given in the text.  The predicted absolute
magnitudes are computed from the relations given by \citet{warn87} and
\citet{warn}, and the quoted uncertainties reflect {\it only} the uncertainty
in the inclination.  The absolute magnitudes at maximum light is computed
from the apparent magnitude using the geometrically-based distance from
Table 3 (excluding prior magnitude information).  The uncertainty reflects
only the uncertainty in distance; other less quantifiable uncertainties not
taken into account include the appropriateness of the figure used for 
$V_{\rm max}$.
}
\end{deluxetable}

\clearpage
\begin{figure}
\plotone{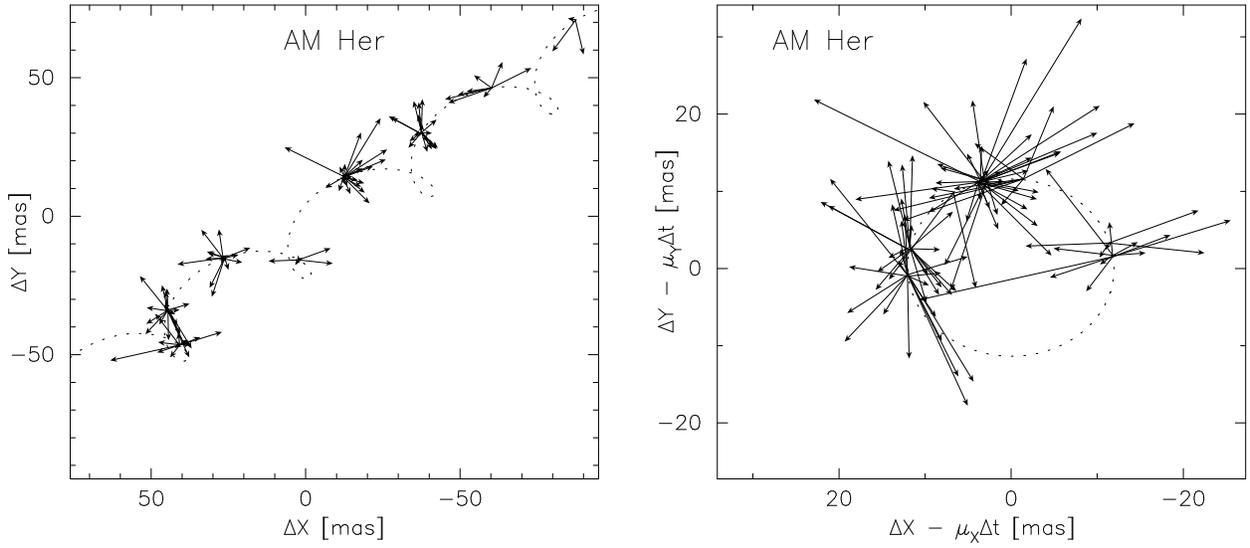}
\caption{Fits to the positions of AM Her referred to its mean position.
The tip of each arrow gives the position in a single image, and the tail
is the position predicted by the fitted parallax, proper motion, and zero
point.  The left panel shows the trajectory across the sky, while the 
right panel shows the data and fit in a reference frame moving with the 
fitted proper motion, leaving only the parallax displacement.  The parallactic
ellipse is nearly circular because AM Her lies near the ecliptic pole.
}
\end{figure}
\clearpage
\begin{figure}
\plotone{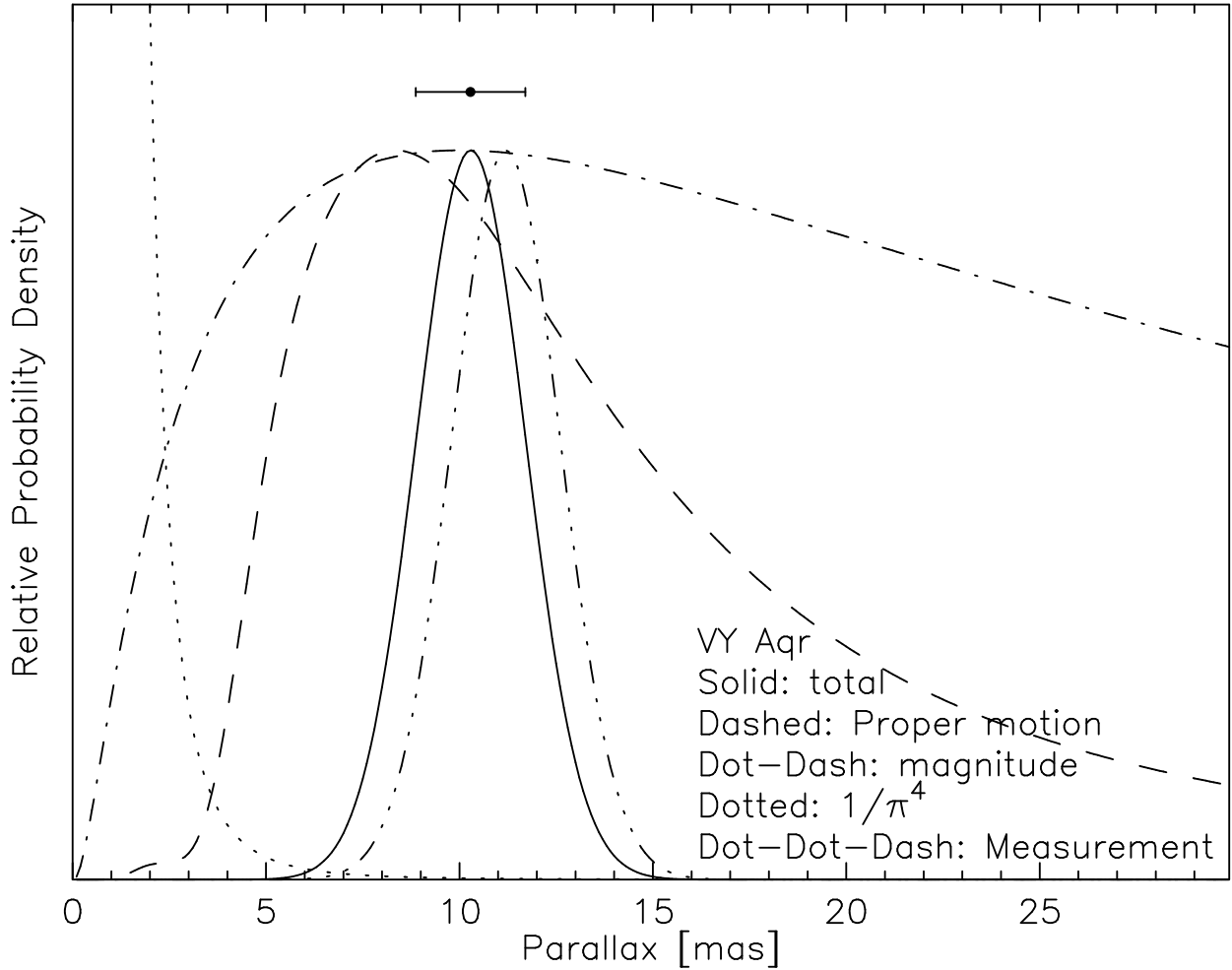}
\caption{Illustration of the various contributions to the Bayesian 
distance-estimation procedure described in the text, for VY Aqr.  
The solid curve gives the net result, and the error bar above
it gives the equivalent 1$\sigma$ confidence interval.  In this 
case the parallax measurement (dot-dot-dash curve) is precise 
enough that the other factors serve only to shift the result to slightly
smaller parallaxes.  The relative normalization of the various 
factors is arbitrary.
}
\end{figure}
\clearpage
\begin{figure}
\plotone{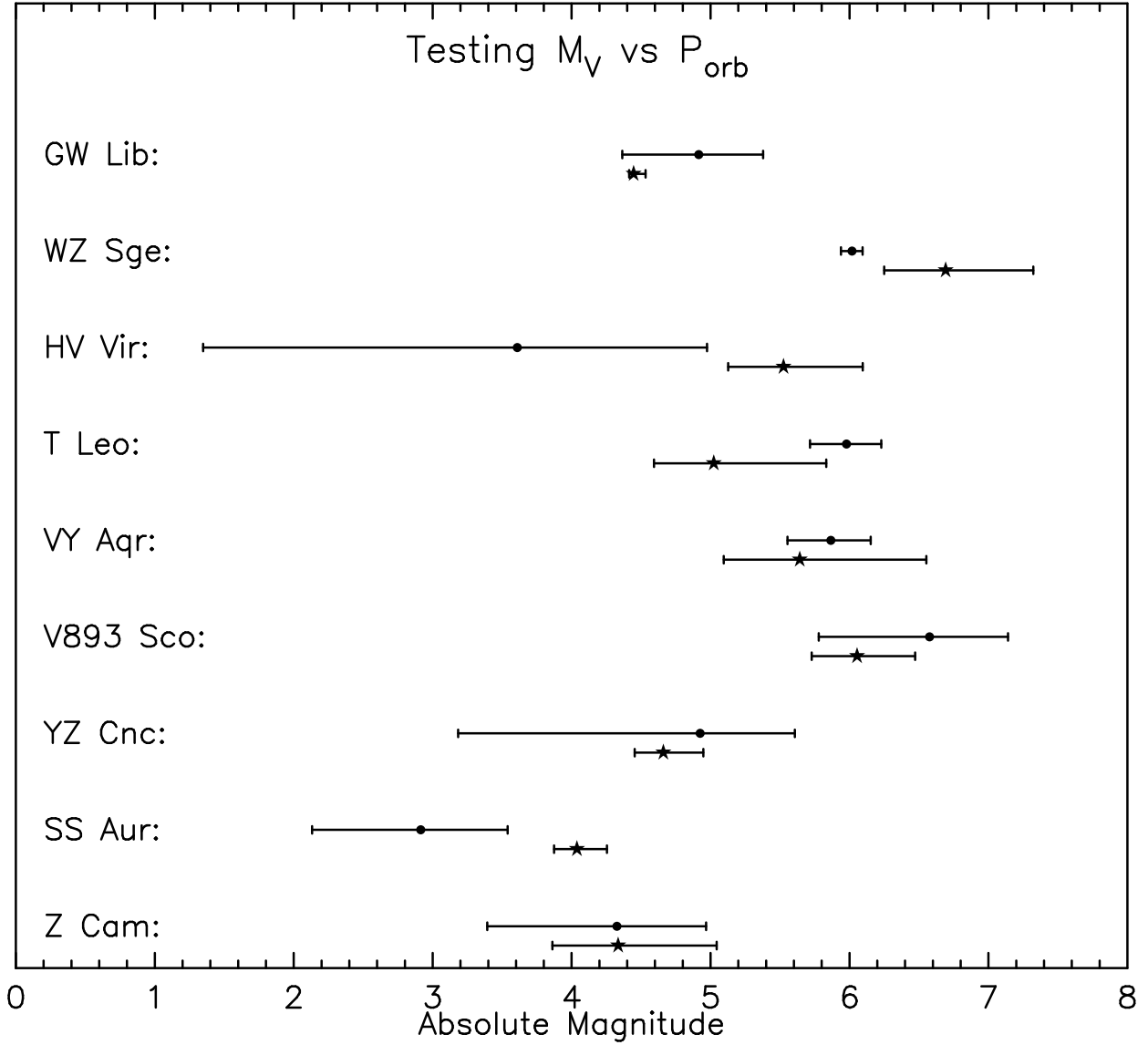}
\caption{Graphical presentation of the empirical and predicted 
absolute $V$ magnitude from
Table 4.  The stars are arranged in order of increasing period.
The top error bar in each set (round dot) is the empirical 
value formed from the apparent $V$ magnitude at maximum light
and the distance inferred from the parallax and proper motion 
(only); the lower error bar (star) is the value predicted by the 
$M_{V}$ (max) - $P_{\rm orb}$ relation.  See the comments
to Table 4. 
}
\end{figure}


\clearpage

\begin{thebibliography}

% \bibitem[Baraffe \& Kolb(2000)]{bk00} Baraffe, I., \& Kolb., U. 2000, \mnras,
% 318, 354

\bibitem[Bailey(1981)]{bailey81} Bailey, J.\ 1981, \mnras, 197, 
31 

\bibitem[Berriman(1987)]{berriman87} Berriman, G.\ 1987, \aaps, 
68, 41 

\bibitem[Berriman, Szkody, \& Capps(1985)]{bsc} Berriman, 
G., Szkody, P., \& Capps, R.~W.\ 1985, \mnras, 217, 327 

\bibitem[Bertin \& Arnouts(1996)]{bertin96} Bertin, E.~\& 
Arnouts, S.\ 1996, \aaps, 117, 393 

\bibitem[Bessell(1990)]{bessell} Bessell, M.~S.\ 1990, \pasp, 
102, 1181 

\bibitem[Beuermann et al.(1999)]{beuermann99} Beuermann, K.,
Baraffe, I., \& Hauschildt, P. 1999, \aap, 348, 524

% EF Eri substellar 2ndary paper, requires distance.
\bibitem[Beuermann et al.(2000)]{beuermanneferi} Beuermann, K., 
Wheatley, P., Ramsay, G., Euchner, F., \& G{\" a}nsicke, B.~T.\ 2000, \aap, 
354, L49 

\bibitem[Bruch(1987)]{bruch87} Bruch, A.\ 1987, \aap, 172, 187 

\bibitem[Bruch, Steiner, \& Gneiding(2000)]{bruchv893} Bruch, A.,
Steiner, J.~E., \& Gneiding, C.~D.\ 2000, \pasp, 112, 237

\bibitem[Cash(1979)]{cash79} Cash, W.\ 1979, \apj, 228, 939 

\bibitem[Dahn et al.(2002)]{dahn02} Dahn, C.~C.~et al.\ 2002, 
\aj, 124, 1170 

\bibitem[Dhillon et al.(2000)]{dhillon00} Dhillon, V. S., 
Littlefair, S. P., Howell, S. B., Ciardi, D. R., Harrop-Allin, 
M. K., and Marsh, T. R.  2000, \mnras, 314, 826

\bibitem[Duerbeck(1999)]{duerbeck99} Duerbeck, H.~W.\ 1999, 
Informational Bulletin on Variable Stars, 4731, 1 

\bibitem[Friend et al.(1990)]{friend90} Friend, M. T., Martin, J. S., 
Smith, R. C., and Jones, D. H. P.\ 1990, \mnras, 246, 637

\bibitem[Gubler \& Tytler(1998)]{gublertytler} Gubler, J.~\& Tytler, 
D.\ 1998, \pasp, 110, 738 

% \bibitem[Hanson(1979)]{hanson79} Hanson, R.~B.\ 1979, \mnras, 
% 186, 875 

% Harrison et al.'s 1999 breakthrough paper
\bibitem[Harrison et al.(1999)]{harrison99} Harrison, T.~E., 
McNamara, B.~J., Szkody, P., McArthur, B.~E., Benedict, G.~F., Klemola, 
A.~R., \& Gilliland, R.~L.\ 1999, \apjl, 515, L93 

% Harrison et al. -- 2ndaries aren't on MS. (duh).
\bibitem[Harrison et al.(2000)]{harrison00} Harrison, T.~E., McNamara, B.~J., 
Szkody, P., \& Gilliland, R.~L.\ 2000, \aj, 120, 2649 

% Harrison et al 2003 - EF Eri boondoggle.
\bibitem[Harrison et al.(2003)]{harrison03} Harrison, T.~E., 
Howell, S.~B., Huber, M.~E., Osborne, H.~L., Holtzman, J.~A., Cash, J.~L., 
\& Gelino, D.~M.\ 2003, \aj, 125, 2609 

\bibitem[Harrison et al.(2003a)]{harrison03a} Harrison, T.~E., 
Johnson, J.~J., McArthur, B.~E., Benedict, G.~F., Szkody, P., 
Howell, S.~B., \& Gelino, D.~M.\ 2003, \aj, in preparation.

% ah her inclination is 46 pm 3 degr.
\bibitem[Horne, Wade, \& Szkody(1986)]{horneahher86} Horne, K.,
Wade, R.~A., \& Szkody, P.\ 1986, \mnras, 219, 791

% T leo multiwavelength study with crappy S/O light curve -> V = 10
\bibitem[Howell et al.(1999)]{howelltleo} Howell, S.~B., Ciardi, 
D.~R., Szkody, P., van Paradijs, J., Kuulkers, E., Cash, J., Sirk, M., \& 
Long, K.~S.\ 1999, \pasp, 111, 342    

% U Gem parallaxes  ... must get from ILL to be sure.
\bibitem[Kamper(1979)]{kamper79} Kamper, K.~W.\ 1979, IAU 
Colloq.~53: White Dwarfs and Variable Degenerate Stars, 494 

% Kato T Leo superoutburst gets good v mag for Super.
\bibitem[Kato(1997)]{katotleo} Kato, T.\ 1997, \pasj, 49, 583 

% Kato V893 sco recovery.
\bibitem[Kato et al.(1998)]{katov893i} Kato, T., Haseda, K., 
Takamizawa, K., Kazarovets, E.~V., \& Samus, N.~N.\ 1998, Informational 
Bulletin on Variable Stars, 4585, 1 

% Kato V893 sco magn.
\bibitem[Kato, Matsumoto, \& Uemura(2002)]{katov893ii} Kato, T., 
Matsumoto, K., \& Uemura, M.\ 2002, Informational Bulletin on Variable 
Stars, 5262, 1 

% Lick Northern Proper Motion catalog description.
\bibitem[Klemola, Jones, \& Hanson(1987)]{npm} Klemola, 
A.~R., Jones, B.~F., \& Hanson, R.~B.\ 1987, \aj, 94, 501 

\bibitem[Kolb \& Stehle(1996)]{kolbstehle} Kolb, U.~\& Stehle, R.\ 
1996, \mnras, 282, 1454 

% mean abs mag of the U Gem variables
\bibitem[Kraft \& Luyten(1965)]{kraftluyten65} Kraft, R.~P.~\& 
Luyten, W.~J.\ 1965, \apj, 142, 1041 

\bibitem[Kraft, Krzeminski, \& Mumford(1969)]{kraftzcam} Kraft, R. P., 
Krzeminski, W., and Mumford, G. S. 1969, \apj, 158, 589 

\bibitem[Landolt(1992)]{landolt92} Landolt, A. U. 1992, AJ, 104, 340

\bibitem[Leibowitz et al.(1994)]{leib94} Leibowitz, E. M., Mendelson, H.,
Bruch, A., Duerbeck, H. W., \& Seitter, W. C. 1994, \apj, 421, 771

\bibitem[Loredo(1992)]{loredo92} Loredo, T. J. 1992, in Feigelson, E. D.,
\& Babu, G. J. (ed.) Statistical Challenges in Modern Astronomy,
New York:Springer Verlag, p. 275

\bibitem[Lutz \& Kelker(1973)]{lutzkelker} Lutz, T.~E.~\& Kelker, 
D.~H.\ 1973, \pasp, 85, 573 

\bibitem[Marsh(1999)]{marsh99} Marsh, T.~R.\ 1999, \mnras, 304, 
443 

\bibitem[McArthur et al.(1999)]{mcarthurrwtri} McArthur, B.~E.~et 
al.\ 1999, \apjl, 520, L59 

\bibitem[McArthur et al.(2001)]{mcarthurtvcol} McArthur, B.~E.~et 
al.\ 2001, \apj, 560, 907 

\bibitem[Mennickent \& Diaz(2002)]{mennickent02} Mennickent, R. E.,
and Diaz, M. P. 2002, \mnras, 336, 767

\bibitem[Mihalas \& Binney(1981)]{mihalbinn} Mihalas, D., \& Binney, J.,
Galactic Astronomy, 2nd ed., (Freeman:San Francisco)

\bibitem[Monet \& Dahn(1983)]{monetdahn83} Monet, D.~G.~\& Dahn, 
C.~C.\ 1983, \aj, 88, 1489 

\bibitem[Monet et al.(1992)]{monet92} Monet, D.~G., Dahn, 
C.~C., Vrba, F.~J., Harris, H.~C., Pier, J.~R., Luginbuhl, C.~B., \& Ables, 
H.~D.\ 1992, \aj, 103, 638 (USNO92)

\bibitem[Monet et al.(1996)]{mon96} Monet, D. et al. 1996,
USNO-A2.0, (U. S. Naval Observatory, Washington, DC)

\bibitem[Monet et al.(2003)]{mon03} Monet, D.~G.~et al.\ 
2003, \aj, 125, 984 

\bibitem[North et al.(2002)]{north02} North, R.~C., Marsh, 
T.~R., Kolb, U., Dhillon, V.~S., \& Moran, C.~K.~J.\ 2002, \mnras, 337, 
1215 

% AAVSO study of Z Cam.
\bibitem[Oppenheimer, Kenyon, \& Mattei(1998)]{oppenheimeraavso} 
Oppenheimer, B.~D., Kenyon, S.~J., \& Mattei, J.~A.\ 1998, \aj, 115, 1175 

\bibitem[Patterson (1979)]{patterson79} Patterson, J. 1979, \aj, 84, 804

\bibitem[Patterson et al.(1993)]{pattvy93} Patterson, J., Bond, 
H.~E., Grauer, A.~D., Shafter, A.~W., \& Mattei, J.~A.\ 1993, \pasp, 105, 
69 

\bibitem[Patterson et al.(2002)]{pattwz01ob} Patterson, J.~et al.\ 
2002, \pasp, 114, 721 

\bibitem[Pickles(1998)]{pickles} Pickles, A.~J.\ 1998, \pasp, 
110, 863 

\bibitem[Rosenzweig et al.(2000)]{rosensu} Rosenzweig, P., 
Mattei, J.~A., Kafka, S., Turner, G.~W., \& Honeycutt, R.~K.\ 2000, \pasp, 
112, 632 

\bibitem[Shafter \& Szkody(1984)]{shafterszkody84} Shafter, A. M., 
and Szkody, P. 1984, ApJ, 276, 305

% SS Aur orbital period ... 
%\bibitem[Shafter \& Harkness(1986)]{shafterharkness86} Shafter, A. M.,
%and Harkness, R. P. 1988, AJ, 92, 658

\bibitem[Shafter \& Hessman(1988)]{shafterhessman88} Shafter, A. W.,
and Hessman, F. V. 1988, \aj, 95, 178


\bibitem[Sion et al.(1995)]{sionwz} Sion, E.~M., Cheng, F.~H., 
Long, K.~S., Szkody, P., Gilliland, R.~L., Huang, M., \& Hubeny, I.\ 1995, 
\apj, 439, 957 

\bibitem[Smak(1993)]{smak93} Smak, J.\ 1993, Acta Astronomica, 
43, 101 

\bibitem[Smith(1987a)]{smith87a} Smith, H.\ 1987, \aap, 171, 336 

\bibitem[Smith(1987b)]{smith87b} Smith, H.\ 1987, \aap, 171, 342 

\bibitem[Spogli et al.(2002)]{spogli} Spogli, C., Fiorucci, M., Tosti, G., 
Ciprini, S., Nucciarelli, G.,
Macchia, E., and Monacelli, G. 2002, IBVS No. 5276

\bibitem[Sproats, Howell, \& Mason(1996)]{shm} Sproats, 
L.~N., Howell, S.~B., \& Mason, K.~O.\ 1996, \mnras, 282, 1211 

\bibitem[Spruit \& Rutten(1998)]{spruit} Spruit, H.~C.~\& 
Rutten, R.~G.~M.\ 1998, \mnras, 299, 768 

\bibitem[Stetson(1987)]{stetsondao} Stetson, P.~B.\ 1987, \pasp, 
99, 191 

\bibitem[Szkody et al.(2002a)]{szkodyhv} Szkody, P., Gaensicke, B. T., 
Sion, E. M., and Howell, S. B. 2002, \apj, 574, 950 

\bibitem[Szkody et al.(2002b)]{szkodygw} 
Szkody, P., G{\" a}nsicke, B.~T., Howell, S.~B., \& Sion, E.~M.\ 2002, 
\apjl, 575, L79 

% Z Cam
\bibitem[Szkody \& Wade(1981)]{szkodywade81} Szkody, P., and Wade, R. A. 
1981, \apj, 251, 201

% \bibitem[Dummy & Fake(1899)]{dummyfake} Dummy, I. M. A., and Fake, P. D. Q. 1899,
% apj, 2, 3

\bibitem[Thorstensen \& Ringwald(1995)]{thorzcam} Thorstensen, J. R, and Ringwald, 
F. A. 1995, IBVS, No.~4249

\bibitem[Thorstensen et al.(1996)]{tpst} Thorstensen, J. R., Patterson, J., 
Thomas, G., \& Shambrook, A. 1996, \pasp, 108, 73

\bibitem[Thorstensen(1997)]{ccahvz} Thorstensen, J. R. 1997, \pasp, 109, 1241

\bibitem[Thorstensen \& Taylor(1997)]{uvvyv1504} Thorstensen, J. R., \& Taylor,
C. J. 1997, \pasp, 109, 1359

\bibitem[Thorstensen et al.(2002)]{dnshort} Thorstensen, J. R.,
Patterson, J., Kemp, J., \& Vennes, S. 2002, \pasp, 114, 1108

\bibitem[Thorstensen \& Fenton(2003)]{thor03} Thorstensen, J. R., \& 
Fenton, W. H. 2003, \pasp, 115, 37

\bibitem[Thorstensen \& Kirkpatrick(2003)]{thorkirk03} Thorstensen, J. R., \&
Kirkpatrick, J. D. 2003, \pasp, in press.

\bibitem[Thorstensen, Wade, \& Oke(1986)]{thorwade86} Thorstensen, J. R.,
Wade, R. A., and Oke, J. B. 1986, \apj, 309, 721

\bibitem[van Paradijs, Augusteijn, \& Stehle(1996)]{vanparadijs96} 
van Paradijs, J., Augusteijn, T., \& Stehle, R.\ 1996, \aap, 312, 93 

\bibitem[Wallace(1994)]{wallace} Wallace, P.~T.\ 1994, ASP 
Conf.~Ser.~ 61: Astronomical Data Analysis Software and Systems III, 3, 481 

\bibitem[Warner(1987)]{warn87} Warner, B.\ 1987, \mnras, 227, 23 

% Warner book, cited in intro.
\bibitem[Warner(1995)]{warn} Warner, B. 1995, 
Cataclysmic Variables (Cambridge University Press)

% Quest for red secondaries ... AM Her distance.
\bibitem[Young \& Schneider(1981)]{youngschneider81} Young, P., 
and Schneider, D. P. 1981, \apj, 247, 960




\end{thebibliography}
\end{document}